\documentclass[preprint,3p,twocolumn]{elsarticle}
\usepackage{geometry}
\usepackage{fancyhdr}

\biboptions{numbers,sort&compress}
\usepackage{graphicx}
\usepackage{amsmath}
\usepackage{amssymb}
\usepackage{verbatim}
\usepackage{wasysym}
\usepackage{longtable}
\usepackage{url}
\newcommand{\dd}{\,\textrm{d}}
\newcommand{\Pec}{{\rm Pe}}
\newcommand{\Rey}{{\rm Re}}
\newcommand{\Ray}{{\rm Ra}}
\newcommand{\Nus}{{\rm Nu}}
\newcommand{\Pra}{{\rm Pr}}
\newcommand{\Str}{{\rm St}}
\newcommand{\Sch}{{\rm Sc}}
\newcommand{\uu}{{\bf u}}
\newcommand{\vv}{{\bf v}}
\newcommand{\gr}{{\bf g}}
\newcommand{\ii}{{\underline i}}
\newcommand{\jj}{{\underline j}}
\renewcommand{\arraystretch}{1.1}
\journal{Computational Materials Science}
\begin{document}
\begin{frontmatter}
 \title{Multiscale dendritic needle network model of \\alloy solidification with fluid flow}
 \author{D. Tourret$^{1,2}$$^*$}
 \author{M.M. Francois$^2$}
 \author{A.J. Clarke$^{2,3}$}
 \address{$^1$IMDEA Materials Institute, Getafe, 28906 Madrid, Spain}
 \address{$^2$Los Alamos National Laboratory, Los Alamos, NM 87545, USA}
 \address{$^3$George S. Ansell Department of Metallurgical and Materials Engineering, Colorado School Mines, Golden, CO 80401, USA}
\cortext[cor1]{Corresponding author (damien.tourret@imdea.org)}
%
\begin{abstract}
We present a mathematical formulation of a multiscale model for solidification with convective flow in the liquid phase.
The model is an extension of the dendritic needle network approach for crystal growth in a binary alloy.
We propose a simple numerical implementation based on finite differences and step-wise approximations of parabolic dendritic branches of arbitrary orientation.
Results of the two-dimensional model are verified against reference benchmark solutions for steady, unsteady, and buoyant flow, as well as steady-state dendritic growth in the diffusive regime.
Simulations of equiaxed growth under forced flow yield dendrite tip velocities within 10\% of quantitative phase-field results from the literature.
Finally, we perform illustrative simulations of polycrystalline solidification using physical parameters for an aluminum-10wt\%\,copper alloy.
Resulting microstructures show notable differences when taking into account natural buoyancy in comparison to a purely diffusive transport regime.
The resulting model opens new avenues for computationally and quantitatively investigating the influence of fluid flow and gravity-induced buoyancy upon the selection of dendritic microstructures.
Further ongoing developments include an equivalent formulation for directional solidification conditions and the implementation of the model in three dimensions, which is critical for quantitative comparison to experimental measurements.
\end{abstract}
\end{frontmatter}
\thispagestyle{fancy}
%
\section{Introduction}
\label{sec:intro}

Dendrites are the most common morphology found in as-solidified metallic alloy microstructures~\cite{Langer80, TrivediKurz94}.
Because the morphological features of these microstructures directly affect the properties of structural materials, understanding and predicting dendritic growth is key to the control and design of technological solidification processes, such as casting, welding, and additive manufacturing.

Dendritic patterns result from the underlying crystalline symmetry of the solid, combined with local thermal and chemical conditions in the surrounding fluid.
Thus, dendritic growth involves coupled mechanisms across a broad scale range: from the atomic structure of the solid-liquid interface to the macroscopic transport of heat and species in the fluid.
Classical theories of dendritic growth have first focused on two fundamental phenomena at play, namely,  diffusion and capillarity~\cite{Ivantsov47, BarbieriLanger89, BenAmarBrener93}. 
In comparison, mechanisms of microstructure selection in presence of fluid flow are less understood.

Yet, crystal growth yields local solute and temperature inhomogeneities that, under the effect of gravity, lead to substantial buoyant convection in the fluid.
The influence of gravity-induced convection on solidification microstructure has been acknowledged and studied for decades~\cite{Mehrabian70, NguyenThi89, Dupouy89}.
Its occurrence makes it nearly impossible to perform solidification experiments of bulk samples under homogeneous conditions~\cite{Jamgotchian01}.
For over 20 years, this has provided a strong motivation for experiments in reduced gravity~\cite{Glicksman94, NguyenThi05, NguyenThi17}.

At the macroscopic scale, liquid advection is responsible for macrosegregation of solute and the formation of solidification defects such as of highly segregated channels, also known as freckles~\cite{Flemings74, Beckermann02, HeinrichPoirier04, Shevchenko13}.
At the scale of the dendritic microstructure, one important effect of advection is to break the intrinsic symmetry of crystal growth present in the diffusive regime.
Depending upon the fluid thermophysical properties, e.g. its thermal expansion, or upon the relative weight of solute and solvent, the flow pattern may also become unstable~\cite{Shevchenko13, Mathiesen06, Steinbach09, Gibbs16}.

Regarding dendritic growth in the presence of fluid flow, some fundamental problems have been addressed analytically, such as the selection of dendrite tip morphology~\cite{BouissouPelce89}.  
Analytical solutions often use the approximation of a diffusive boundary layer (also sometimes referred to as a ``stagnant film'') surrounding the dendrite~\cite{CantorVogel77, Ananth91, Sekerka95, LiBeckermann02, Gandin03}.
Beyond the level of complexity of a single dendritic tip, computational methods become necessary to investigate the interactions between fluid flow and crystal growth. 
Computational models have been proposed that consist of explicitly tracking the location of the solid-liquid interface (see, e.g., \cite{Udaykumar03, AlRawahiTryggvason04, ZhaoHeinrichPoirier05}).
Alleviating the high computational cost of explicit interface tracking, most computational studies over the past couple of decades have used the phase-field method~\cite{Steinbach09, Beckermann99, Tong01, Jeong01, Jeong03, Lu05, Rojas15}.
Still, even combining the most advanced state-of-the-art numerical methods and hardware, simulations remain limited to the scale of a handful of dendritic grains~\cite{Sakane18, Takaki18}.

Because of these limitations, several scale-bridging approaches have been proposed to simulate dendritic growth at scales larger than those accessible to phase-field.
These models include continuum volume-averaged approaches~\cite{Beckermann02, HeinrichPoirier04, NiBeckermann91, WangBeckermann96, WuLudwig09_1}, models based on dynamics of average dendritic grain envelopes~\cite{SteinbachBeckermannEtAl99, Souhar16}, and approaches coupling cellular automata with finite elements~\cite{RappazGandin93, GandinRappaz94}, finite differences~\cite{WangLeeMcLean03}, or Lattice Boltzmann methods~\cite{Jelinek2014}.
Granular models, with approximate grain shapes based on Voronoi space tessellation, have also been used to study the occurrence of solidification defects resulting from the coupling between grain growth and fluid flow, such as porosities and hot cracking~\cite{Vernede2006, Sistaninia2012}.
Yet, all these models do not resolve the transient interactions between individual dendritic branches that are crucial to the inner grain dendritic microstructure selection.
In order to bridge the scale gap between phase-field and coarse-grained models, we recently proposed a multiscale Dendritic Needle Network (DNN) approach~\cite{dnn2d,dnn3d}.
The method is suited to modeling the solidification of concentrated alloys, typically forming at low solute supersaturation, when dendrites form hierarchical tree-like structures with several generations of needle-like branches.

The major computational advantage of the DNN method resides in the fact that it retains quantitative predictions for a numerical space discretization of the same order or even larger than dendrite tip radii.
While this is still much finer than what is used in volume-averaged models at the scale of whole cast ingots, this is one order of magnitude coarser than what is required for quantitative phase-field calculations~\cite{echebarria04,shibuta15}.
Requiring ten times fewer grid points (i.e. ten times fewer operations) in each spatial direction and allowing the use of a higher time step (e.g. up to 100 times if using an explicit time scheme), DNN simulations can be orders of magnitude faster than using phase-field.
While the model does not predict complex morphological details of the solid-liquid interface, it enables computationally-efficient simulations of dendritic arrays at the larger scale of heat and mass transport.

For purely diffusive transport, we have already presented the detailed derivation of the model in two dimensions (2D)~\cite{dnn2d} and three dimensions (3D)~\cite{dnn3d}.
The resulting model was thoroughly verified by quantitative comparisons to exact analytical solutions for transient and steady state growth, to phase-field calculations results, and also validated against measurements from directional solidification experiments~\cite{dnn3d,mcwasp,jom,mertens}.

The following article presents a first two-dimensional derivation of the DNN model that includes convective flow in the liquid phase.
As a first step, we only treat the case of solute-driven growth of a binary alloy at a fixed, homogeneous temperature, with negligible solute diffusion in the solid.
We give a description of the model (Sec.~\ref{sec:model}) and its current numerical implementation (Sec.~\ref{sec:implem}).
We test the fidelity of the model and separately verify the calculations of fluid flow and dendritic growth (Sec.~\ref{sec:valid}).
Then, we compare quantitative predictions of the model coupling fluid flow and dendritic growth against phase-field results (Sec.~\ref{sec:DNNvsPF}).
Finally, we illustrate the potential of the model with simulations of polycrystalline solidification with and without gravity-induced buoyancy, using realistic parameters for a metallic alloy.

\section{Model}
\label{sec:model}

\subsection{Dendritic needle network}

The DNN model~\cite{dnn2d,dnn3d} aims at simulating solidification at low solute supersaturation, i.e. at low P\'eclet number $\Pec\equiv R V/(2D)$, with $R$ and $V$ the dendrite tip radius and velocity respectively, and $D$ the solute diffusivity in the liquid.
In these conditions, the scale of the dendritic tip radius is much lower than the scale of solute transport in the liquid, such that one can describe a dendritic grain as a hierarchical network of sharp branches, and conservation equations can be derived at different length scales, including at an intermediate scale much larger than $R$, but much smaller than the diffusion length $l_D=D/V$.
The instantaneous growth conditions of each needle-like branch, namely $R(t)$ and $V(t)$, can thus be calculated by combining a solute conservation condition at this intermediate scale that gives the time evolution of $R V^2$ (in 2D) with a standard microscopic solvability condition that is established at the scale of the tip radius and prescribes the value of $R^2V$~\cite{BarbieriLanger89,BenAmarBrener93}.

\subsubsection{Sharp-interface problem}
\label{sec:sharpintpb}

We consider the growth of a binary alloy of nominal solute concentration $c_\infty$ at a given temperature $T=T_0$ lower than its liquidus temperature $T_L$.
At the scale of the microstructure, crystal growth in a liquid of homogeneous temperature is usually a reasonable approximation for metallic alloys in which heat diffuses several orders of magnitude faster than solute. 
The evolution of the solid-liquid interface can be modeled with the sharp interface problem consisting of: (i) a description of transport within bulk phases, (ii) a statement of conservation at the solid-liquid interface, and (iii) a condition for equilibrium, or departure from equilibrium, at the interface. 
Since details are provided in Refs~\cite{dnn2d,dnn3d}, here we directly start from the dimensionless form of the corresponding equations. 
For clarity, an exhaustive list of notations is given in \ref{app:symbols}.

We introduce the dimensionless form of the solute concentration, i.e. the solute supersaturation field
\begin{align}
\label{eq:defU}
U \equiv \frac{c_0-c}{(1-k)c_0},
\end{align}
where $c$ is the solute concentration, $c_0$ is the liquid equilibrium concentration at the temperature $T_0$, and $k$ is the interface solute partition coefficient (here considered a constant).
Neglecting solute transport in the solid, in the vicinity of the interface the solute concentration in the liquid locally obeys the diffusion equation
\begin{align}
\label{eq:DiffU}
\partial_tU &= D\nabla^2U .
\end{align}
Under the aforementioned assumptions and neglecting the capillary correction to the concentration jump across the interface, typically small, the Stefan condition for the conservation of solute at the interface, which relates the normal gradient of $U$ at the interface, $\partial_n U|_i$, to the interface normal velocity, $v_n$, can be approximated as
\begin{align}
\label{eq:StefanU}
v_n &= D \partial_n U |_i  .
\end{align}
Finally, if one neglects the kinetic undercooling of the interface, local equilibrium can be expressed using the Gibbs-Thomson condition
\begin{align}
\label{eq:GT}
U_i &= d_0 f_\gamma(\bar\theta) \kappa ,
\end{align}
where $\kappa$ is the interface curvature.
The solute capillarity length $d_0$ is given by 
\begin{align}
\label{eq:capillength}
d_0 = \frac{\Gamma_{sl} }{|m|(1-k)c_0},
\end{align}
where $m$ is the alloy liquidus slope ($m<0$) and $\Gamma_{sl}$ is the Gibbs-Thomson coefficient of the interface.
The anisotropy function $f_\gamma(\bar\theta)$ describes the dependence of the interface stiffness $\gamma(\bar\theta)+\partial_{\theta\theta}\gamma(\bar\theta)=\gamma_0 f_\gamma(\bar\theta)$ upon the interface orientation $\bar\theta$, with $\gamma(\bar\theta)$ the orientation-dependent excess free energy of the interface, and $\gamma_0$ its value averaged over all orientations in a (100) plane~\cite{Haxhimali06,Dantzig13}.
The sharp interface problem is thus defined by Eqs.~\eqref{eq:DiffU}-\eqref{eq:GT}, which are combined with an imposed supersaturation $U=\Omega$ far away from the interface, with 
\begin{align}
\label{}
\label{eq:supersaturation}
\Omega \equiv \frac{c_0-c_\infty}{(1-k)c_0}.
\end{align}

\subsubsection{Microscopic solvability at the dendrite tip scale}

\begin{figure}[b]
\centering
\includegraphics[width=3.in]{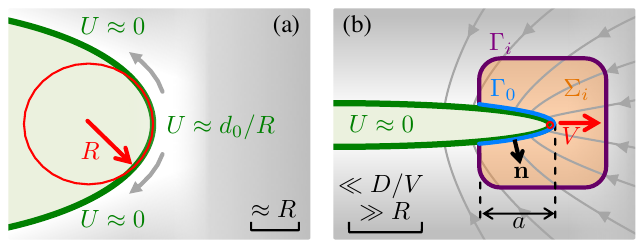}
\caption{
\label{fig:DNN} 
The tip radius $R(t)$ and velocity $V(t)$ of each needle-like dendritic branch in the DNN model is established by combining two conditions at distinct length scales:
(a) a solvability condition at the scale of the tip radius $R$ and 
(b) a conservation condition at an intermediate scale larger than the tip radius $R$, but smaller than the scale of transport in the liquid (typically the diffusion length $l_D=D/V$).
}
\end{figure}

At the scale of the dendrite tip radius $R$ (Fig.~\ref{fig:DNN}a), the free-boundary problem defined by Eqs.~\eqref{eq:DiffU}-\eqref{eq:GT} only has a solution if the interface energy and stiffness are anisotropic~\cite{BarbieriLanger89,BenAmarBrener93}.
The product $R^2 V$ is thus given by
\begin{align} 
\label{eq:R2V}
R^2 V = \frac{2 D d_0}{\sigma}
\end{align}
with a selection parameter $\sigma$ that depends solely on the magnitude of crystalline anisotropy. 

This solvability condition was extensively verified by phase-field simulations, e.g. in Refs~\cite{KarmaRappel98,ProvatasEtAl98,PlappKarma00}.
While Eq.~\eqref{eq:R2V} was initially proposed for a steady-state shape-preserving dendrite, PF calculations have shown that the value of $R^2 V$ reaches a constant value early during the transient development of a dendrite tip~\cite{PlappKarma00}.
Therefore, the DNN model makes use of the relation \eqref{eq:R2V} both in the early-stage and in the steady-state growth regimes. 

Within the range of parameters accessible to the phase-field method, simulations have shown the constancy of $\sigma$ for a fluid velocity up to about one order of magnitude higher than the tip velocity~\cite{Tong01, Jeong03, Lu05}.
This is consistent with solvability theory with fluid flow, which states that the value of $\sigma$ only changes when the fluid velocity is much higher than the tip growth velocity~\cite{BouissouPelce89}.
With $u$ the fluid velocity, $\sigma$ is expected to remain constant as long as $(d_0 u)/(R V)\ll 1$, i.e. as long as $\Pec\sigma u/V\ll 1$~\cite{BouissouPelce89, Lu05}.
This means that the fluid velocity needs to be larger than the tip velocity by a factor $1/(\sigma\Pec)$ to significantly affect the tip selection.
The tip selection parameter $\sigma$ is of the order of $10^{-2}$, and the current model is dedicated to solidification at low P\'eclet number $\Pec\ll1$.
Therefore, the condition for the constancy of $\sigma$ will practically always be satisfied, and we will use a constant $\sigma$ value, uniquely given by the anisotropy of the solid-liquid interface.

\subsubsection{Solute conservation at intermediate scale}
\label{sec:intermedscale}

At a scale much larger than its tip radius $R$ (Fig.~\ref{fig:DNN}b), a dendritic branch appears essentially sharp and curvature effects can be neglected, such that along a needle-like branch, we have 
\begin{equation}
\label{eq:Uequal0}
U = U_i \approx 0,
\end{equation}
similarly as for a flat interface.

At this scale, we can assume the quasi-stationary growth of a shape-preserving parabola of equation $y_i(x)=\sqrt{2R(x_t-x)}$ at a velocity $V$ in the $x+$ direction.
The conservation equation~\eqref{eq:StefanU} over a length $a$ behind the tip, i.e. along a contour $\Gamma_0$ between $x_t$ and $x_t-a$ as in Fig.~\ref{fig:DNN}b, yields
\begin{align} 
D \int_{\Gamma_0}\frac{\partial u}{\partial n}\dd\Gamma_0 &= 
 \int_{-y_a}^{y_a} \hspace{-10pt}{V \dd y} \\ 
&= 2V\sqrt{2R a} ,
\end{align}
with $y_a\equiv y_i(x_t-a)=\sqrt{2R a}$. 
Thus, introducing the dimensionless flux intensity factor (FIF)
\begin{align} 
\label{eq:newFIF2D}
\mathcal F \equiv \frac{1}{4\sqrt{a/d_0}} \; \int_{\Gamma_0}\big( \partial_n U \big) \dd\Gamma_0 ,
\end{align}
the product $R^2V$ is given by
\begin{align} 
\label{eq:RV2}
R V^2 = \frac{2 D^2 \mathcal F^2}{d_0} .
\end{align}

Furthermore, in the immediate vicinity of the tip at a scale much lower than $D/V$, one can consider Laplace equation in a moving frame of velocity $V$ 
\begin{align} 
\label{eq:Laplace}
D \nabla^2 U = -V \partial_x U .
\end{align}
Thus, one way to estimate the time-dependent $\mathcal F(t)$ is to integrate $\nabla^2 U$ over a surface $\Sigma_i$ bounded by the contours $\Gamma_i$ and $\Gamma_0$ (see Fig.~\ref{fig:DNN}b) and to use the divergence theorem to write
\begin{align} 
\label{eq:intgcontourthick}
\int_{\Gamma_0}\big(\partial_n U\big)\dd\Gamma_0 = \int_{\Gamma_i}\big( \partial_n U \big) \dd\Gamma_i + \frac{V}{D} \iint_{\Sigma_i} \big( \partial_x U \big) \dd\Sigma_i ~,
\end{align}
where the normal $\mathbf{n}$ is taken pointing outward of the closed contour $(\Gamma_i+\Gamma_0)$ along $\Gamma_i$ and outward of the dendrite, i.e. inward the contour, along $\Gamma_0$.
Hence, $\mathcal F(t)$ can be calculated from Eq.~\eqref{eq:newFIF2D}, where the integral along $\Gamma_0$ is given by Eq.~\eqref{eq:intgcontourthick} using any chosen integration contour $\Gamma_i$ joining the needle at $x=x_t-a$.

\subsubsection{Conservation equations in the liquid phase at the macroscopic scale}

While transport is predominantly diffusive in the immediate vicinity of the solid-liquid interface, we also consider fluid flow within the bulk liquid phase.
In this first version of the model, the crystals are motionless, such that there is no velocity in the solid phase and the transport equations presented below are only solved in the liquid phase.

The liquid is assumed incompressible and Newtonian, such that the conservation of mass yields the incompressibility condition
\begin{align}
\label{eq:mass}
\nabla \cdot \uu =\;&0 ,
\end{align}
and the conservation of momentum is given by the Navier-Stokes equations, here in their conservative form (see e.g.~\cite{michels,gdn})
\begin{align}
\label{eq:momentum}
\varrho\big[ \partial_t \uu + \nabla\cdot(\uu\,\uu) \big] &= {\bf F_V} - \nabla p + \eta \nabla^2 \uu ,
\end{align}
where $\uu$ is the fluid velocity field, $\varrho$ is the fluid density, $p$ is the pressure field, $\eta$ is the dynamic fluid viscosity (here considered constant), and ${\bf F_V}$ represents volume forces.

For a buoyant flow under natural convection, density variations can be ignored except in terms involving the gravity $\gr$, i.e using the Boussinesq approximation commonly assumed in solidification problems (see e.g. \cite{HeinrichPoirier04, Steinbach09}).
Within the considered isothermal conditions, density variations are solely due to solute composition differences, such that volume forces are
\begin{align}
\label{eq:Boussinesq}
{\bf F_V} = \varrho\,\gr = \varrho_0\left[ 1 - \beta_c(c-c_0)\right]\gr,
\end{align}
where the expansion coefficient is assumed constant and estimated in the vicinity of the reference state $c=c_0$  at which $\varrho=\varrho_0$ with
\begin{align}
\label{eq:expcoeff}
\beta_c &\equiv-\left.\frac{1}{\varrho_0}\frac{\partial\varrho}{\partial c}\right|_{c=c_0}.
\end{align}

Assuming an ideal dilute solution (i.e. following Fick's 1$^{\rm st}$ law) and no source term, the transport of solute in the bulk liquid follows the standard advection-diffusion equation 
\begin{align}
\label{eq:solute}
\partial_t c + \nabla\cdot(\uu\,c) = \nabla \left(D \nabla c\right),
\end{align}
where the equation was simplified by a factor $\varrho_0$.

\subsection{Scaling}
\label{sec:scaling}

For consistency with nondimensional scaling applied in Refs~\cite{dnn2d,dnn3d}, space and time are scaled using values of the tip radius, $R_s$, and velocity, $V_s$, of an isolated free dendrite in a steady growth regime.

The theoretical steady state of a free dendrite can be found by combining the microscopic solvability condition with a condition on the steady-state tip P\'eclet number, $\Pec \equiv R_sV_s/2D$.
A natural choice for this additional relation, which we use here, is the analytical Ivantsov solution~\cite{Ivantsov47}
\begin{align} 
\label{eq:Iv2D}
{\rm Iv}(\Pec) = \sqrt{\pi \Pec} \, \exp(\Pec) \, \textrm{erfc}\left(\sqrt{\Pec}\right),
\end{align}
which describes the steady-state diffusion around a shape-preserving parabola growing at a given supersaturation $\Omega={\rm Iv}(\Pec)$.

Nondimensional time is noted $\tau = t V_s/R_s$, and the scaled equations describing the evolution of $\tilde R\equiv R(t)/R_s$ and $\tilde V\equiv V(t)/V_s$ are similar to those given in given in Ref.~\cite{dnn3d} (Sec.~3.1 therein)
\begin{align} 
\label{eq:R2V_scaled}
\tilde R^2 \tilde V &= 1 \\
\label{eq:RV2_scaled}
\tilde R \tilde V^2 &= 2 \alpha^2 \tilde{\mathcal F}^2 ,
\end{align}
with 
\begin{equation} 
\label{eq:FIF_scaled}
\tilde{\mathcal F} =  \frac{1}{4\sqrt{\tilde a}} \; \int_{\Gamma_0}\frac{\partial U}{\partial n}\dd\Gamma_0 ,
\end{equation}
where $\alpha \equiv D/(R_s V_s)$ is the dimensionless solute diffusivity, and $\tilde a\equiv a/R_s$.
The scaled conservation equations in the liquid reduce to 
\begin{align}
\label{eq:mass_scaled}
\nabla \cdot \vv &=0 \\
\label{eq:momentum_scaled}
\partial_\tau \vv + \nabla\cdot(\vv\,\vv) &= \chi \nabla^2 \vv - \nabla \psi + \left[ 1 + \lambda_c U \right] \tilde\gr \\
\label{eq:solute_scaled}
\partial_\tau U + \nabla\cdot(\vv\,U) &= \alpha \nabla^2 U .
\end{align}
The nondimensional unknowns are the solute field $U$, the fluid velocity $\vv \equiv \uu/V_s$, and the pressure field $\psi \equiv p/(\varrho_0 V_s^2)$.
In addition to the scaled diffusivity $\alpha$, reduced parameters are the kinematic viscosity $\chi \equiv \nu/(R_s V_s)$ with $\nu=\eta/\varrho_0$, the gravity $\tilde\gr \equiv R_s\gr/V_s^2$, and the solutal expansion coefficient $\lambda_c \equiv (1-k)c_0\beta_c$. 
(Also see list of notations in \ref{app:symbols}.)

This scaled formulation reveals the classical nondimensional numbers that govern the dynamics of the system~\cite{michels,gdn}, namely a Reynolds number
$\Rey^* = (V_sR_s)/\nu = 1/\chi$,
the 
Schmidt number $\Sch = \nu/D = \chi/\alpha$,
and the Rayleigh number $\Ray = \lambda_c|\tilde\gr|/(\alpha\chi)$.
The star superscript in $\Rey^*$ denotes that it stands for a Reynolds number relative to the crystal growth, as it is scaled with the tip growth velocity $V_s$, while we simply note $\Rey = (R_s u_0)/\nu$ the more usual Reynolds number scaled with respect to a characteristic fluid velocity $u_0$.
The steady state dendrite tip P\'eclet number can also be expressed as $\Pec \equiv (R_sV_s)/(2D) = 1/(2\alpha)=\Sch\,\Rey^*/2$.

In the remainder of this article, coordinates and their corresponding partial derivatives are scaled with respect to $R_s$ and $R_s/V_s$, even though we simply use notations $x$, $y$, and $t$ for the sake of clarity.

\section{Numerical implementation}
\label{sec:implem}

Developed in two dimensions, Eqs~\eqref{eq:mass_scaled}-\eqref{eq:solute_scaled} are 
\begin{align}
\label{eq:incomp2D}
\partial_x u+\partial_y v =~& 0 \\
\label{eq:u2D}
\partial_t u + \partial_x (u^2) + \partial_y (uv) =~& \chi \left( \partial_{xx} u + \partial_{yy} u  \right) \nonumber\\ &
 - \partial_x \psi + f_x(U) \\
\label{eq:v2D}
\partial_t v + \partial_x (uv) + \partial_y (v^2) =~& \chi \left( \partial_{xx} v + \partial_{yy} v \right) \nonumber\\ &
 - \partial_y \psi + f_y(U)  \\
\label{eq:U2D}
\partial_t U + \partial_x (u\,U) + \partial_y (v\,U) =~& \alpha \left( \partial_{xx} U + \partial_{yy} U \right) 
\end{align}
with
\begin{align}
\label{eq:bodyf}
f_\mu(U) = \left[ 1 + \lambda_c U \right] \tilde g_\mu ,
\end{align}
where $\tilde g_\mu$ denotes the scaled gravity component along the axis $\mu\in\{x,y\}$. 

We combine Eqs~\eqref{eq:incomp2D}-\eqref{eq:bodyf} with the DNN growth kinetics Eqs~\eqref{eq:R2V_scaled}-\eqref{eq:FIF_scaled}.
The scale separation requirement for rigorously deriving the DNN growth equations are met, as long as the scale of the tip radius, $R$, and that of the tip flux integration domain, $a$, remain much smaller than the predominantly diffusive boundary layer (or ``stagnant film'') surrounding the dendrite.
This is typically the case unless convection is extremely strong~\cite{BouissouPelce89, clarke2017}.
Thus, as we are mostly interested in natural gravity-induced buoyancy, we can (i) use DNN growth equations established in the diffusive case, and (ii) use standard numerical methods for relatively low Reynolds number.

We choose to keep numerical methods relatively simple, at the expense of computing time and numerical accuracy. 
The resolution of Navier-Stokes Eqs~\eqref{eq:incomp2D}-\eqref{eq:U2D} is done similarly as in Ref.~\cite{gdn}.
Hence, here we only summarize the main numerical techniques, and the reader is referred to Ref.~\cite{gdn} (Chapters 3 and 9) for further details.

\subsection{Time stepping}
\label{sec:timestepping}

We solve the advection-diffusion equation \eqref{eq:U2D} explicitly, i.e. in discrete form
\begin{align}
\label{eq:explicitU}
&\frac{U^{(n+1)}-U^{(n)}}{\Delta t}  = \nonumber\\&
\Big[ \alpha \partial_{xx} U + \alpha \partial_{yy} U - \partial_x (u\,U) - \partial_y(v\,U) \Big]^{(n)}
\end{align}
where superscripts $^{(n)}$ and $^{(n+1)}$ denote current and next time steps respectively, and $\Delta t$ is the time step.

The incompressibility Eq.~\eqref{eq:mass} is not a time evolution equation but an algebraic condition, which we can treat separately using a standard projection approach~\cite{Chorin68}.
Following this method, we first evolve the momentum equations \eqref{eq:u2D}-\eqref{eq:v2D} neglecting pressure terms, and then project the solution on the subspace of divergence-free velocity fields.

Thus, for the first step, we discretize the pressure-less momentum conservation Eq.~\eqref{eq:momentum_scaled} explicitly as
\begin{align}
\label{eq:projector}
\frac{\vv^{(*)}-\vv^{(n)}}{\Delta t} &= \Big[ \chi \nabla^2 \vv - \nabla\cdot(\vv\,\vv) 
+ \left( 1 + \lambda_c U \right) \tilde\gr \Big]^{(n)} 
\end{align}
where the superscript $^{(*)}$ denotes intermediate estimated values of the velocity field.

In the next step, the relation between the pressure field $\psi$ and the continuity equation is described by the Poisson equation~\cite{Chorin68}
\begin{align}
\label{eq:poisson}
\nabla^2 \psi^{(n+1)}= \frac{1}{\Delta t}\Big[ \nabla\cdot \vv^{(*)} \Big]
\end{align}
which has to be solved using boundary conditions $\nabla\psi\cdot{\mathbf n}=0$.
We solve Eq.~\eqref{eq:poisson} iteratively with a standard successive over relaxation (SOR) method, with a relaxation parameter $\omega_{\rm sor}$~\cite{Frankel50, Young54}. 
We use a checkerboard (also known as {\it red-black}) ordering, because it is adapted to parallelization on graphic processing units, which does not provide control over the order of parallel threads within kernels (see Sec.~\ref{sec:parallel}).

In the last step of the projection method, knowing $\psi^{(n+1)}$, the velocity field is corrected as
\begin{align}
\label{eq:p_correc}
\vv^{(n+1)} = \vv^{(*)}-\Delta t \; \nabla \psi^{(n+1)}.
\end{align}

In order to ensure numerical stability, we use a variable time step $\Delta t$ that is adapted following
\begin{align}
\label{eq:adaptdt}
\Delta t = K_{\Delta t} \times \min\left\{ \Delta t_{\rm max} ~,~ \frac{h}{\max\{u,v\}}\right\}
\end{align}
where $K_{\Delta t}$ is a user-input safety factor, $h$ is the spatial grid step size, and the stability limit for the explicit time discretization of diffusion and viscosity terms is 
\begin{align}
\label{eq:stabdt}
\Delta t_{\rm max} = \begin{cases}
h^2/(4\,\alpha\,\Sch) &\text{if $\Sch>1$},\\
h^2/(4\,\alpha) &\text{otherwise}.
\end{cases}
\end{align}
                                                                                                
\subsection{Space discretization}
\label{sec:spacediscr}

\begin{figure}[!b]
\centering
\includegraphics[width=3.in]{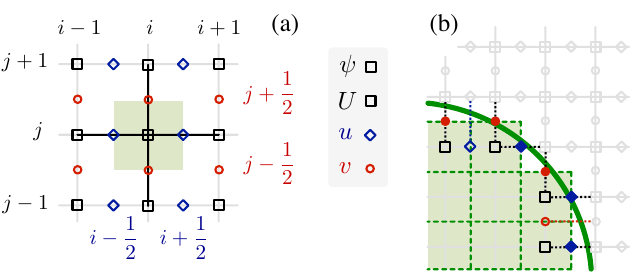}
\caption{
\label{fig:grid} 
(a) Staggered grid with shifted velocity components $u$ and $v$ and centered fields $\psi$ and $U$ along the finite difference grid. 
(b) Illustration of a curved solid-liquid interface (thick solid green line) represented as steps (thin dashed green line), showing the location of grid points where a specific boundary condition is imposed directly (full symbols) or indirectly (open symbols).
}
\end{figure}

We spatially discretize the equations using finite differences on a structured grid of square elements of side $h$ along both $x$ and $y$ directions. 
To avoid nonphysical solutions, such as checkerboard pressure distributions due to incompressibility, we use a staggered grid~\cite{michels,gdn}.
Fig.~\ref{fig:grid}a shows the five-point stencil centered on a grid point $(i,j)$, with its equivalent finite volume cell as shaded green background.
The $x$ component of the velocity field $u$ ($\diamond$) and the $y$ component of the velocity field $v$ ($\circ$) are expressed at the center of a link between two grid points in their respective directions, i.e. shifted by $h/2$ in $x$ or $y$ respectively, while $\psi$ and $U$ are expressed at the location of grid points ($\square$). 

We use centered differences for diffusive terms in Eqs~\eqref{eq:u2D}-\eqref{eq:U2D}.
For convective terms, we use a weighted average between centered differences and a donor-cell scheme~\cite{michels, gdn}, with an upwind parameter $0\leq\omega_{\rm up}\leq1$ that is zero for a centered difference scheme and one for a donor-cell scheme. 

Although less precise, this discretization allows more flexibility on the selection of the grid size $h$. 
The latter can be typically chosen $h\approx R$, as long as $h\ll D/V$~\cite{dnn2d,dnn3d}, hence providing a good compromise between accurate predictions and stability for a relatively wide range of convective conditions at low to moderate Reynolds numbers.

Discretized terms are fully developed in \ref{app:discr}: Eq.~\eqref{eq:U2D} in \ref{app:discrsolute}, Eqs~\eqref{eq:u2D}-\eqref{eq:v2D} excluding pressure terms in \ref{app:discrpredict}, and the SOR iteration of the pressure Poisson Eq.\eqref{eq:poisson} in \ref{app:discrpoisson}.

\subsection{Boundary conditions}
\label{sec:bc}

Boundary conditions are applied at the center between grid points, i.e. at the equivalent boundary of square control volumes centered on the grid points (green shaded cell in Fig.~\ref{fig:grid}a). 

For a staggered field stored along those boundaries, a Dirichlet condition can be directly imposed. 
When the field is not defined at the location of the boundary, we use a linear combination of neighboring values. 
For instance, the condition $u_{i+\frac{1}{2},N+\frac{1}{2}}=u_{bc}$ is imposed as
\begin{align}
u_{i+\frac{1}{2},N+1} = 2u_{bc}-u_{i+\frac{1}{2},N}.
\end{align}
Similarly, to impose a Neumann condition, such as $\partial_y\psi=0$ for the Poisson equation \eqref{eq:poisson} along the same $j=N+1/2$ boundary, one simply applies 
\begin{align}
\psi_{i,N+1} =\psi_{i,N}.
\end{align}
Boundary conditions on the intermediate field $\vv^{(*)}$ are similar to those on $\vv^{(n+1)}$.
Boundary conditions for $\psi^{(n+1)}$ in Eq.~\eqref{eq:poisson} are $\nabla\psi\cdot{\mathbf n}=0$~\cite{gdn,Chorin68}.

External boundaries and internal solid-liquid boundaries are treated similarly.
Along the latter, we impose a Dirichlet boundary condition on both the solute field $U=0$ (equilibrium condition) and the velocity field, $u=0$ and $v=0$ (no-slip condition).
These boundaries are only evolving due to crystal growth, since grain motion is not currently taken into account.

Curved boundaries are treated as steps, as illustrated in Fig.~\ref{fig:grid}b.
Each grid point ($\square$) within the theoretical interface (thick solid green line) is considered as the center of a square fully-solid cell.
The resulting step-wise solid appears as green background with the solid cell tiling in dashed green lines in Fig.~\ref{fig:grid}b.
Grid points where a boundary condition is directly imposed are in full symbols.
Grid points where a boundary condition is imposed through a linear combination of neighbor values are in open symbols.
Grid points in light gray are treated as regular liquid points.
Another illustration of the step-wise approximation for a parabola appears in Fig.~\ref{fig:fif}, discussed later.

Both external and internal boundaries are stored in a mask integer field $\phi$ that is used to filter fluid cells and boundary cells.
The latter includes several types of boundaries, depending on the number and location of neighbor fluid cells (see Ref.~\cite{gdn}).
Fluid cells surrounded by boundary cells in two opposite directions (i.e. $x-$ and $x+$, or $y-$ and $y+$) are treated as boundary cells.

\subsection{Dendritic branching and growth}
\label{sec:fif}

\subsubsection{New branch addition}
\label{sec:branching}

Similarly as in Refs~\cite{dnn2d,dnn3d}, when dendritic sidebranching is enabled, we generate new branches on the sides of a needle at a distance $l_{sb}$ behind the tip of the growing needle, every time the latter grows by a distance $l_{sb}$.
In order to mimic natural fluctuations~\cite{PietersLanger86, CouderEtAl05}, sidebranching events are randomized and decoupled from one another by generating a random fluctuation $\delta l_{sb}$ within a range $[-\Delta l_{sb}/2;\Delta l_{sb}/2]$ when creating a new branch, and then setting the distance for generating the next sidebranch at $l_{sb}=\big[L_{sb}+\delta l_{sb}\big]R_s$.
The average sidebranching frequency $L_{sb}$ and the amplitude of its fluctuation $\Delta l_{sb}$ are user input parameters chosen to match typical experimental measurements, e.g. a sidebranching distance from the tip between 5 and 10 times the tip radius~\cite{CouderEtAl05, MelendezBeckermann12}.
While this representation of sidebranching frequency could be improved (see e.g. \cite{Sturz16}), it has limited influence on final selected microstructures at the scale of whole dendritic arrays, as long as sidebranches are created often enough to be in growth competition with one another~\cite{dnn2d}.

The initial length of a sidebranch generated on the side of a needle of current tip radius $R$ is set to $\sqrt{2R l_{sb}}+R$, i.e. one $R$ longer than the parabolic half-width of the parent needle at the location of the new branch.
Note that we only treat sidebranching for fourfold symmetry crystals in this article. 

\subsubsection{Individual branch growth}
\label{sec:growth}

Each dendritic branch grows following Eqs~\eqref{eq:R2V} and \eqref{eq:RV2}, or their nondimensional equivalent Eqs.~\eqref{eq:R2V_scaled}-\eqref{eq:RV2_scaled}.
Besides the unknown $R(t)$ and $V(t)$, the only time-dependent term in those equations is $\mathcal F(t)$, defined in Eq.~\eqref{eq:newFIF2D}.
It is calculated using Eq.~\eqref{eq:intgcontourthick} with an outer contour $\Gamma_i$ intersecting the needle at a distance $a$ behind the tip (Fig.~\ref{fig:DNN}b).

\begin{figure}[!b]
\centering
\includegraphics[width=2.2in]{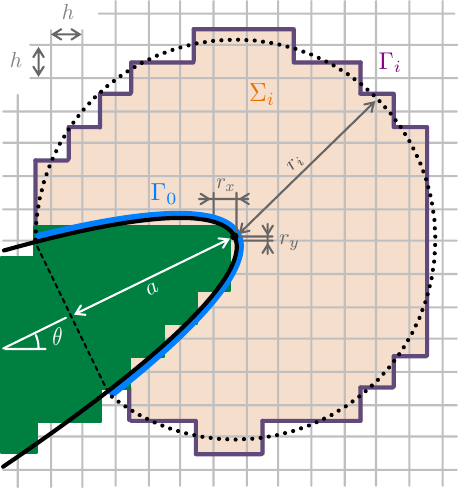}
\caption{
\label{fig:fif} 
Vicinity of a needle tip titled by an angle $\theta$, showing the theoretical parabolic shape of the dendrite tip (solid black line) and the circular integration contour (dotted black line) of radius $r_i$ intersecting the needle at a distance $a$ behind the tip location.
Approximate step-wise descriptions of the solid domain appear in solid green shading.
The integration surface $\Sigma_i$ delimited by the integration contour $\Gamma_i$ is in light orange shading.
}
\end{figure}

We consider the growth of needles in any direction, thus allowing sixfold (hexagonal) grains (see e.g. Fig.~\ref{fig:rotated}) and polycrystalline microstructures (see Sec.~\ref{sec:Polycrystal} and Ref.~\cite{dnncet_tms}).
For a good compromise between simplicity and generality in the calculation of $\mathcal F$, we use a circular integration contour $\Gamma_i$ of radius $r_i$ and centered on the parabolic needle tip, as illustrated in Fig.~\ref{fig:fif}.
The integration circle and the parabolic needle (and thus the grid cells captured as part of either, as described in Section~\ref{sec:bc} and Fig.~\ref{fig:grid}b) both advance at each time step.
Using this contour, the value of $a$ is directly given by $\sqrt{R^2+r_i^2}-R$ (see \ref{app:acirc}).

The circular contour is numerically approximated with steps, similarly as the solid domain boundaries (see Sec.~\ref{sec:bc}).
Each grid point within the circular domain is considered at the center of an integration cell (orange shaded background in Fig.~\ref{fig:fif}).
Practically, we define a local mask function $q_{i,j}$ that is equal to 1 inside the contour and 0 outside.
Then, integrating in the liquid region in the vicinity of the needle tip, the surface integral over $\Sigma_i$ in Eq.~\eqref{eq:intgcontourthick} is calculated as  
\begin{align}
\iint_{\Sigma_i} \big( \partial_x U \big) \dd\Sigma_i = \frac{h}{2}\sum_{i,j}\Big\{q_{i,j} &\big[(U_{i+1,j}-U_{i-1,j})\cos\theta \nonumber\\ &
+ (U_{i,j+1}-U_{i,j-1})\sin\theta \big] \Big\},
\end{align}
with $\theta$ the growth direction of the needle (see Fig.~\ref{fig:fif}).
The contour integral along $\Gamma_i$ in Eq.~\eqref{eq:intgcontourthick} is also integrated in the vicinity of the needle tip as  
\begin{align}
\int_{\Gamma_i}\big( \partial_n U \big) \dd\Gamma_i = \sum_{i,j}&\Big\{(q_{i,j}-q_{i+1,j})\times (U_{i,j}-U_{i+1,j}) \nonumber\\ &
+ (q_{i,j}-q_{i-1,j})\times (U_{i,j}-U_{i-1,j}) \nonumber\\ &
+ (q_{i,j}-q_{i,j+1})\times (U_{i,j}-U_{i,j+1}) \nonumber\\ &
+ (q_{i,j}-q_{i,j-1})\times (U_{i,j}-U_{i,j-1}) \Big\}.
\end{align}

In some cases (e.g. Section~\ref{sec:ivantsov}), we track the position of the most advanced needle tip, and periodically shift the fields such that the dendritic front remains at a fixed $x$ location in the grid.

Finally, in the case of growth competition among several branches, when a needle slows down and $V\to0$, the constancy of $R^2V$ from Eq.~\eqref{eq:R2V} might cause numerical instabilities.
One way to avoid them is to truncate the parabola, for instance bounding the maximal half-width of each needle to a given $r_{max}$.
As shown in Ref.~\cite{dnn3d}, this approximation yields reasonably accurate results, as long as the truncation occurs behind the tip at a length greater than $a$, i.e. as long as the needle is not truncated within the integration contour $\Gamma_i$.

\subsection{Data structure and parallel algorithm}
\label{sec:parallel}

We solve the problem numerically using Graphics Processing Units (GPUs) with the CUDA parallel computing platform from Nvidia\textsuperscript\textregistered~\cite{cuda}.

Arrays are allocated on the GPU (device) for single-precision real number fields $u^{(n)}$, $u^{(*)}$, $v^{(n)}$, $v^{(*)}$, $U^{(n)}$, $U^{(n+1)}$, $\psi^{(n)}$, and $\psi^{(n+1)}$, and the integer solid flag field $\phi^{(n)}$ and $\phi^{(n+1)}$.
Time stepping on the $U$ field is performed by swapping pointer addresses between steps $(n)$ and $(n+1)$ after computing $U^{(n+1)}$.
Velocities $u^{(n+1)}$ and $v^{(n+1)}$ are directly updated in arrays $u^{(n)}$ and $v^{(n)}$.
The pressure field $\psi$ is updated at each SOR iteration within the same array.
Arrays $\psi^{(n+1)}$ and $\phi^{(n+1)}$ are only allocated to facilitate the parallelization of the the shifting of the moving domain, if enabled.
We also allocate a GPU array of a {\it Needle} structure that contains individual needle properties, e.g. origin, direction, length, tip radius and velocity.
We track the number of needles in the simulation at the end of each time step, and if necessary we resize the array to ensure that it is filled between 70\% and 90\% with active needles.

\begin{table*}[!h]
\centering
\caption{
Time stepping algorithm. Bold font indicates steps performed on the GPU. Star superscripts in the {\it Step} column denote optional steps that are only performed if necessary or at a given frequency (e.g. file output).
}%
\center%
\label{tab:algo}%
\begin{tabular}{ l l l l l }%
\hline
Step & Action & Performed on & Parallelization & Eq.\\
\hline
{1} & {Update time step $\Delta t$} & {CPU (host)} &  & {\eqref{eq:adaptdt}}  \\
\bf{2} & \bf{Compute $U^{(n+1)}$} & \bf{GPU (device)} & \bf{Blocks} & {\eqref{eq:explicitU}}\\
\bf{3} & \bf{Compute $u^{(*)}$, $v^{(*)}$} & \bf{GPU} & \bf{Blocks} & {\eqref{eq:projector}} \\
\bf{4} & \bf{Compute $\psi^{(n+1)}$} & \bf{GPU} {(CPU)} & \bf{Blocks} {(Iterative)} & {\eqref{eq:poisson}} \\
{5} & {Reset $||\psi||_2$ and $\max\{u,v\}$ to zero} & {CPU} &  &  \\
{\bf 6} & {\bf Update $u^{(n+1)}$, $v^{(n+1)}$} (also integrating $||\psi||_2$) & \bf{GPU} & \bf{Blocks} & {\eqref{eq:p_correc}} \\
\bf{7} & \bf{Calculate $R$, $V$ for each needle} & \bf{GPU} & \bf{Needles} & {\eqref{eq:R2V}-\eqref{eq:RV2}} \\
{8}$^*$ & {Locate most advanced needle tip} & {CPU} &  &  \\
{\bf 9}$^*$ & \bf{Shift domain} & \bf{GPU} & \bf{Blocks} &  \\
{\bf 10}$^*$ & \bf{Generate sidebranches} & \bf{GPU} & \bf{None} &  \\
{11}$^*$ & {Count needles} & {CPU} &   &  \\
\bf{12} & \bf{Update boundary condition flag $\phi$} & \bf{GPU} & \bf{Blocks} & \\
\bf{13} & \bf{Apply boundary conditions} & \bf{GPU} & \bf{Blocks} &  \\
\bf{14} & \bf{Find $\max\limits_{i,j}\{u,v\}$} & \bf{GPU} & \bf{Reduction} &  \\
{15}$^*$ & {File output} & {CPU} & &  \\
{16}$^*$ & {Resize Needle array} & {CPU} &   &  \\
\hline
\end{tabular}
\end{table*}

One array per field is allocated on the CPU (host) for $u$, $v$, $\psi$, $U$, and $\phi$, and the needle array.
Data copies of fields occur only during the main initialization before the time stepping loop (host to device), and whenever a file output is necessary (device to host).
The needle structure array is copied between host and device when: (i) tracking the position of all needle tips, (ii) counting the total number of needles, (iii) redimensioning of the needle array, and (iv) file output, all of which are performed on the CPU.

The resulting time step loop is summarized in Table~\ref{tab:algo}.
Heavy calculations, in bold font, are performed on the GPU, mostly using a parallelization by {\it Blocks}, i.e. spatially splitting the domain into blocks of typical size $16\times32$ grid points.
The needle growth step (7), during which $R$ and $V$ are calculated for each individual needle, is parallelized using one thread per needle (labeled {\it Needles} in Table~\ref{tab:algo}).
The sidebranching step (10) is not parallelized, i.e. running a single GPU thread, to ensure that the addition of new needles in the array occurs in an orderly manner.
(It could be parallelized with appropriate thread synchronization.)
In step (14), in order to find $\max\limits_{i,j}\{u,v\}$, we use a standard reduction algorithm~\cite{reduc}. 
Before this step, data of arrays $u$ and $v$ was previously reduced to a single array containing $\max\{u_{i,j},v_{i,j}\}$ during step (6).

In step (4), the pressure Poisson Eq.~\eqref{eq:poisson} is solved with an iterative SOR loop monitored for convergence at the CPU level.
Each iteration includes two GPU kernel calls, iterating Eq.~\eqref{eq:iterSOR} over grid points with even and odd values of $(i+j)$, respectively.
Using an even number of grid points in both $x$ and $y$, this results in the two-step checkerboard pattern of the red-black SOR iteration.
A residual $r_{\rm sor}$ is initialized to zero at the start of the loop, incremented during GPU calculations (see \ref{app:discrpoisson}), and compared to a tolerance $\overline r_{\rm sor}$ on the CPU at the end of the loop.
The residual is calculated for each iteration $(it)$ as 
\begin{align}
\label{eq:residual}
r_{\rm sor} \equiv \frac{\Big|\Big| \nabla^2 \psi^{(it)} - \frac{1}{\Delta t}\left[ \nabla\cdot \vv^{(*)} \right] \Big|\Big|_2}{\big|\big| \psi^{(n)} \big|\big|_2}
\end{align}
with 
\begin{align}
\big|\big| \xi \big|\big|_2\equiv\Bigg[\sum\limits_{i,j} \left(\xi_{i,j}\right)^2 \Bigg]^{1/2}
\end{align}
the $L^2$-norm of a field $\xi$.
The iterative process is stopped when $r_{\rm sor}\leq\overline r_{\rm sor}$ or after a maximum number of iteration $N_{\rm iter}$, typically here with $N_{\rm iter}=100$ (although the loop usually converges in just a few iterations).

Step (12), which updates the solid and boundary mask field $\phi$ consists of three separate GPU kernel calls, parallelized by blocks.
The first one resets the whole array to fluid cells with only external boundaries cells.
The second updates the array with solid cells at the locations of dendrites.
The third identifies edge cells that are fluid but adjacent to at least one solid cell.

Efficiency could be further improved, for instance by reducing the amount of memory copies between host and device by performing steps (8) and (11) on the GPU (e.g. with reduction algorithms~\cite{reduc}), or parallelizing step (10) using advanced thread synchronization.
However, since most of the heavy calculations are performed in parallel on the GPU, this implementation already provides a substantial acceleration.
Using an unsteady fluid flow simulation for $\Rey=100$ (i.e. Fig.~\ref{fig:vonkarman}b of Sec.~\ref{sec:vonkarman}) as a benchmark, we found the GPU-parallelized calculations to be over 10 times faster on a single Nvidia\textsuperscript{\textregistered} Tesla\textsuperscript{\textregistered} K40c than a similar serial calculation using a single core of a Intel\textsuperscript{\textregistered} Xeon\textsuperscript{\textregistered} CPU E5-2660 v2 (2.20~GHz). 

\section{Verification of the numerical implementation}
\label{sec:valid}

Before using the model coupling dendritic growth and fluid flow, we verify the individual components of the current numerical implementation.
First, we test fluid dynamics for a steady flow (Sec.~\ref{sec:steadyflow}), a steady buoyant flow (Sec.~\ref{sec:steadybuoyancy}), and an unsteady flow with a fixed obstacle (Sec.~\ref{sec:vonkarman}).
Then, we test the growth of a dendritic grain in diffusive conditions (Sec.~\ref{sec:isodnn}).

\subsection{Steady flow}
\label{sec:steadyflow}

\begin{figure}[!t]
\centering
\includegraphics[width=3.in]{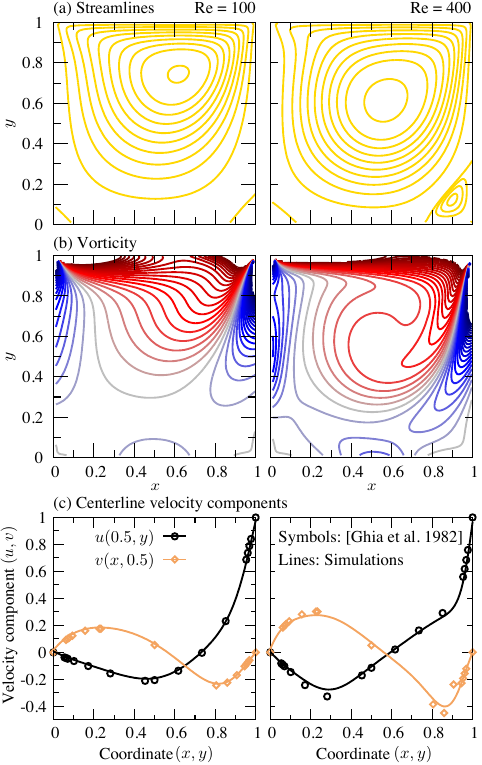}
\caption{
\label{fig:ghia} 
Simulation results for lid-driven cavity steady flow for $\Rey=100$ (left) and $\Rey=400$ (right): (a) Streamlines; (b) Iso-values of the vorticity field; and (c) Velocity component $u(y)$ along the $x=0.5$ centerline and $v(x)$ along the $y=0.5$ centerline (lines) compared to results from Ref.~\cite{Ghia82} (symbols).
Contour plots show similar iso-values for both $\Rey$, with a smaller step for negative values of the stream function to illustrate the recirculation loop in the bottom right corner for $\Rey=400$ (a).
}
\end{figure}

We test the resolution of the steady Navier-Stokes equations for a lid-driven flow in a square cavity~\cite{Ghia82}.
The domain size is $L_x\times L_y=1\times1$ using $128^2$ grid points (including boundaries), i.e. $h\approx0.0079365$.
Boundary conditions are $(u,v)=(0,0)$ along left, bottom, and right boundaries, and $(u,v)=(1,0)$ at the top boundary.
We perform simulations for two Reynolds numbers $\Rey=100$ and 400, respectively for a dimensionless total time of 20 and 100.
Numerical parameters are $K_{\Delta t}=0.6$, $\omega_{\rm up}=0.9$, $\omega_{\rm sor}=1.7$, and $\overline r_{\rm sor}=10^{-4}$.

Figure~\ref{fig:ghia} shows the resulting streamlines (a), vorticity fields (b), and the $u$ and $v$ profiles across a centerline (c).
(See \ref{app:vortstr} for the definition of the stream function and vorticity field.)
The comparison between computed velocity profiles (lines) and reference results by Ghia et al.~\cite{Ghia82} (symbols) provides a verification of the current implementation up to moderate Reynolds numbers, starting to show a small deviation only at $\Rey=400$.

\subsection{Steady buoyant flow}
\label{sec:steadybuoyancy}

\begin{figure}[!t]
\centering
\includegraphics[width=3.in]{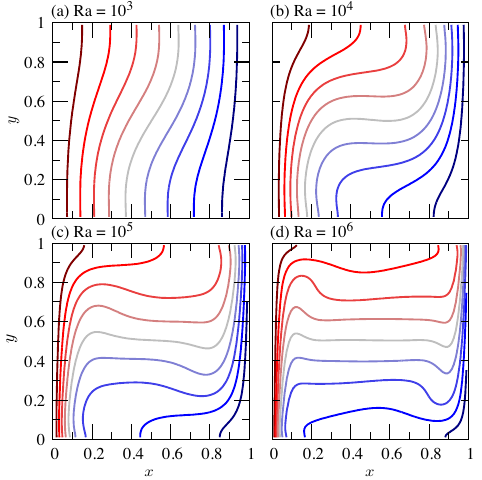}
\caption{
\label{fig:buoyancy} 
Steady solute field from simulations of a square cavity with imposed boundary conditions $U=1$ at $x=0$ and $U=0$ at $x=1$ with step of 0.1, for different Rayleigh numbers $\Ray=10^3$ (a), $10^4$ (b), $10^5$ (c), and $10^6$ (d).
}
\end{figure}

We test buoyancy using the reference benchmark by de Vahl Davis~\cite{deVahlDavis83}.
The reference problem is formulated for heat transport in a square cavity induced by differentially heated walls, but it is similarly applicable to solute transport.

The domain size is $L_x\times L_y=1\times1$ using $94^2$ grid points (including boundaries), hence with $h\approx0.0106$.
Boundary conditions are $(u,v)=(0,0)$ on all boundaries, $\partial U/\partial y=0$ along $y=0$ and $y=1$, $U=1$ at $x=0$, and $U=0$ at $x=1$.
The total dimensionless time is 25, with $\Rey=100$ and $\Sch=0.71$, equivalent to the thermal Prandtl number $\Pra=0.71$ in Ref.~\cite{deVahlDavis83}.
Using a gravity $(g_x,g_y)=(0,-1)$, we set the expansion coefficient $\lambda_c=0.140845$, 1.40845, 14.0845, and 140.845, respectively yielding solutal Rayleigh numbers $\Ray=10^3$, $10^4$, $10^5$, and $10^6$.
Numerical parameters are $K_{\Delta t}=0.5$, $\omega_{\rm up}=0.9$, $\omega_{\rm sor}=1.7$, and $\overline r_{\rm sor}=10^{-3}$.

Computed $U$ fields in Fig.~\ref{fig:buoyancy} compare well with the reference calculation (Fig.~4 of Ref.~\cite{deVahlDavis83}).
Quantitatively, Table~\ref{tab:nusselt} compares the average Nusselt number at the $x=1$ boundary, defined as $\Nus =\left[\int_0^{L_y} \partial U(1,y)/\partial x ~{\rm d}y\right]/L_y$, here with $L_y=1$.
\begin{table}[!t]
\caption{%
Nusselt number at the $x=1$ boundary in Fig.~\ref{fig:buoyancy}.
}%
\center%
\label{tab:nusselt}%
\begin{tabular}{ l l l l l }%
\hline
$\Ray$ & $10^3$ & $10^4$ & $10^5$ & $10^6$ \\ 
\hline
Simulations & 1.122 & 2.250 & 4.545 & 8.832 \\
Reference~\cite{deVahlDavis83} & 1.118 & 2.243 & 4.519 & 8.800 \\
\hline
\end{tabular}
\end{table}
Calculated $\Nus$ values differ by less than $0.6\%$ from that in Ref.~\cite{deVahlDavis83}, thus verifying the current numerical calculation of buoyancy terms.

\subsection{Unsteady flow}
\label{sec:vonkarman}

The final test of the fluid flow simulation is for an unsteady oscillatory flow induced by a circular obstacle, known as a von K\'arm\'an vortex street~\cite{VonKarman04, Williamson96}.
This problem has been extensively studied experimentally (see \cite{Williamson96} and references within).
It was shown that the outflow periodicity can be characterized by its Strouhal number $\Str=f d/ u_i$, which depends uniquely on the Reynolds number~\cite{Rayleigh15}, with $f$ the outflow frequency (i.e. the inverse of its oscillation period), $d$ the obstacle diameter, and $u_i$ the inflow velocity.
Above $\Rey\approx49$ the flow becomes unsteady and enters a laminar vortex shedding regime up to $\Rey\approx180$, when the flow becomes progressively three-dimensional~\cite{Williamson96}.
For $49<\Rey<180$, the outflow periodicity follows a universal law~\cite{Williamson96,Williamson88}
\begin{align}
\label{eq:str_rey}
\Str=\frac{-3.3265}{\Rey}+0.1816+0.00016\,\Rey.
\end{align}

\begin{figure}[!b]
\centering
\includegraphics[width=3.in]{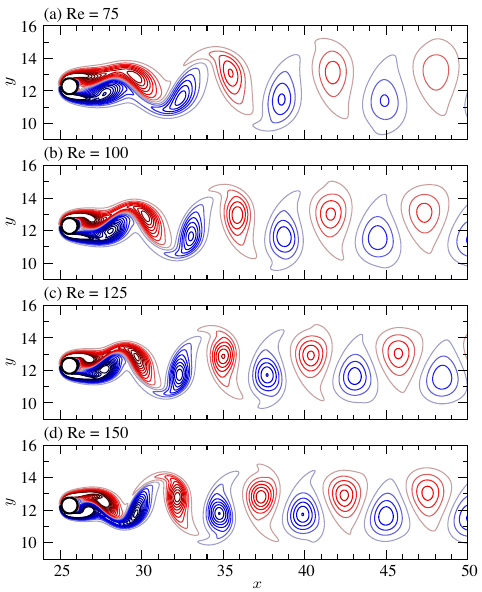}
\caption{
\label{fig:vonkarman} 
Vorticity field of the oscillatory regime in presence of a circular obstacle for an inflow $(u,v)=(1,0)$ at $x=0$ and different $\Rey=75$~(a), 100~(b), 125~(c), and 150~(d).
Contour plots show similar iso-values in all panels, with negative values in blue and positive values in red.
}
\end{figure}

We perform simulations with an obstacle of diameter $d=1$ and an inflow velocity $u_i=1$ in the $x+$ direction.
We use $2560\times 1280$ grid points with a spacing $h=0.02$, hence a domain size $L_x\times L_y$ slightly larger than $50\times25$.
The circular obstacle is at the center of the domain, slightly offset by $d/2$ in the $y$ direction in order to break the symmetry and trigger the flow instability.
The outflow is free at the $x\to\infty$ boundary with $\partial u/\partial x=\partial v/\partial x=0$, and no-slip conditions $(u,v)=(0,0)$ are applied on the top and bottom $y$ boundaries.
Numerical parameters are $K_{\Delta t}=0.5$, $\omega_{\rm up}=0.9$, $\omega_{\rm sor}=1.7$, and $\overline r_{\rm sor}=10^{-3}$.
Total simulated times are 300, 225, 180, and 150, respectively for Reynolds number values $\Rey=75$, 100, 125, and 150.

Fig.~\ref{fig:vonkarman} illustrates the vorticity field (see \ref{app:vortstr}) for $\Rey=75$~(a), 100~(b), 125~(c), and 150~(d). 
\begin{table}[!t]
\caption{%
Strouhal number from simulations compared to experimental law \eqref{eq:str_rey}~\cite{Williamson96,Williamson88}.
}%
\center%
\label{tab:strouhal}%
\begin{tabular}{ l l l l l l }%
\hline
$\Rey$ & 75 & 100 & 125 & 150 \\ 
\hline
Simulations & 0.155 & 0.166 & 0.175 & 0.181 \\
Experiments & 0.149 & 0.164 & 0.175 & 0.183 \\
\hline
\end{tabular}
\end{table}
We calculated the outflow frequency $f$ by fitting a sine function to the $y$ component of the velocity field at the location $(x,y)=(40,L_y/2)$ over the final 50 dimensionless time of the simulation, which corresponds to at least 7 oscillation periods.
Resulting Strouhal numbers appear in Table~\ref{tab:strouhal} compared to Eq.~\eqref{eq:str_rey}.
Predictions fall within a few percents of the expected values, thus verifying the current numerical implementation for unsteady fluid flow.

\subsection{Diffusive dendritic growth}
\label{sec:isodnn}

We now verify the implementation of the model for dendritic growth in the diffusive transport regime.

\subsubsection{Steady isolated free dendrite}
\label{sec:ivantsov}

First, we simulate the steady growth of an isolated free dendrite.
We set the solute supersaturation $\Omega$ and compare the predicted P\'eclet number $\lim\limits_{t\to\infty}\left\{R(t)V(t)/(2D)\right\}$ to the theoretical Ivantsov solution (Eq.~\eqref{eq:Iv2D}).

We use a grid size of $384\times512$ over a time $150 R_s/V_s$ for $\Omega\geq0.2$, $1024\times1536$ over $250 R_s/V_s$ for $0.2>\Omega>0.1$, and $1664\times2240$ over $350 R_s/V_s$ for $\Omega=0.1$.
With $h/R_s=1$ for all simulations, this corresponds to domains of at least $12l_D\times16l_D$, using the scaled diffusion length $l_D/R_s=(D/V_s)/R_s$ given by the Ivantsov Eq.~\eqref{eq:Iv2D}.
We simulate a single needle growing in the $x+$ direction centered in $y$.
The domain is shifted in order to keep the needle tip fixed at 1/4 of the domain size in $x$.
All simulations have $K_{\Delta t}=0.5$, $\omega_{\rm up}=0.9$, $\omega_{\rm sor}=1.7$, and $\overline r_{\rm sor}=10^{-3}$, and no-flux boundary conditions on all boundaries.
Sidebranching is disabled to isolate a unique free dendrite, the thickness of the needles is not bounded (i.e. $r_{max}=\infty$), and the radius for the integration of the flux intensity factor (FIF) is $r_i=5h$ (Fig.~\ref{fig:fif}).

\begin{figure}[!b]
\centering
\includegraphics[width=3.in]{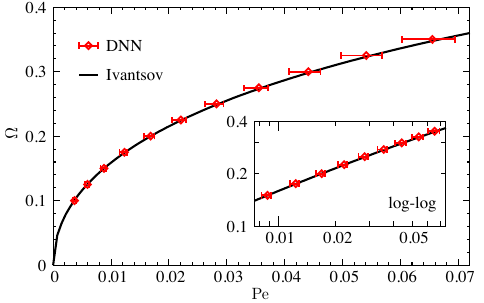}
\caption{
\label{fig:ivantsov} 
DNN predictions of the steady supersaturation $\Omega$ and P\'eclet number $\Pec$ for an isolated dendrite.
Error bars represents the range of numerical oscillations as the needle tip advances between two successive grid points.
The inset shows a log-log representation.
}
\end{figure}

Results in Fig.~\ref{fig:ivantsov} ($\diamond$) show that the steady dendrite growth dynamics is well predicted with the current numerical implementation.
Horizontal error bars represent the range of oscillations of the steady solution as the needle tip crosses one grid element, and symbols are time averaged values, both integrated over the last 10\% of the simulated time.
Oscillations on the P\'eclet number remain within 10\% of the Ivantsov solution for all cases.
The instantaneous P\'eclet number is close to its time average when the tip of the needle is in the middle of two grid points (i.e. here when $r_x/h\approx0.5$ with $r_y=0$ and $\theta=0^\circ$ using notations on Fig.~\ref{fig:fif}). 
In addition to the approximate step-wise description of the needle shape, these oscillations are linked to the scale separation between the fixed grid spacing $h=R_s$ and the diffusion length $l_D=D/V_s$.
Hence, their amplitude is higher for high $\Omega$, since $l_D$ is relatively lower~\cite{dnn3d}, and could be reduced using a lower $h$ for higher $\Omega$.

We observed that predictions in both transient and steady states deviate by at most a few percent, as long as the radius $r_i$ of integration of $\cal F$ is higher than a few grid spacings, e.g. $r_i\geq4h$, and up to about the diffusion length, i.e. $r_i \lesssim D/V$.
We also verified that square and circular integration contours of similar size $r_i$ usually resulted in just a few percent discrepancy in terms of steady state growth velocity (or steady undercooling and supersaturation in directional solidification) for an isolated needle at $\theta=0^\circ$.
Thus, in simulations throughout this paper, we typically use a circular contour as in Fig.~\ref{fig:fif} with $r_i=5R_s$.

\subsubsection{Rotated equiaxed crystal}
\label{}

\begin{figure}[!b]
\centering
\includegraphics[width=3.in]{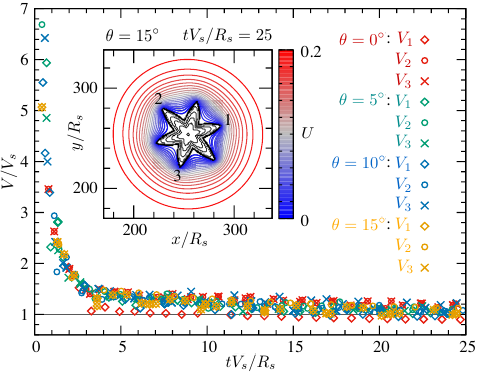}
\caption{
\label{fig:rotated} 
Time evolution of the tip velocity $V$ for needles of hexagonal crystals during isothermal equiaxed growth at a supersaturation $\Omega=0.2$ for different grain orientations $\theta$.
Oscillations as a tip progresses between two successive grip points (see Fig.~\ref{fig:ivantsov}) were filtered out by representing only data points for $|r_x/h-0.5|<0.02$ or  $|r_y/h-0.5|<0.02$ (see notations Fig.~\ref{fig:fif}), i.e. when the needle tip is located in the middle of two grid points in $x$ or $y$.
The inset shows the solute field $U$ from 0 (blue) to 0.2 (red) with steps of 0.01 for $\theta=15^\circ$ at $t=25R_s/V_s$, as well as the evolution grain shape from $t=0$ to $25R_s/V_s$ with step of $5R_s/V_s$ (thin black lines for  $t<25R_s/V_s$ and thick black line for $t=25R_s/V_s$).
}
\end{figure}

We test the numerical implementation for needles not aligned with the numerical grid.
To do so, we simulate the isothermal growth of a hexagonal crystal at $\Omega=0.2$ for different orientations $\theta$, with $\theta$ the angle between a given branch, here labeled 1, and the $x+$ direction (Fig.~\ref{fig:fif}).
A nucleus, composed of six branches of initial length and radius both equal to $R_s$, is set at the center of a domain of dimension $512R_s\times512R_s$, with $h=R_s$. 
Other numerical parameters are similar as in simulations of Fig.~\ref{fig:ivantsov}.

Figure~\ref{fig:rotated} shows the resulting time evolution of the tip velocities $V_1$, $V_2$, and $V_3$ for three branches numbered 1, 2, and 3 in the inset (shown for $\theta=15^\circ$).
Due to the fourfold symmetry of the grid, results are identical for branches of growth direction $\theta$, $-\theta$, $\pi+\theta$, and $\pi-\theta$. 
Thus, results for branches 1 through 3 for $\theta=0^\circ$, 5$^\circ$, 10$^\circ$, and 15$^\circ$ amounts to testing every orientation from 0 to 360$^\circ$ by steps of 5$^\circ$ (even duplicating some, since $V_2\sim V_3$ for $\theta=0^\circ$ and $V_1\sim V_3$ for $\theta=15^\circ$).
Similar evolutions of tip velocities in Figure~\ref{fig:rotated}, as well as the conserved sixfold symmetry of the grain and its surrounding solute field in the inset, provide an additional verification of the numerical implementation of arbitrary grain orientations.
Furthermore, whereas an arbitrary growth orientation may be considered using a rectangular contour~\cite{dnncet_tms}, the current circular contour is more general and consequently more practical from a numerical implementation point of view.

\section{Solidification under convective flow}
\label{sec:DNNvsPF}

\subsection{Equiaxed growth in a forced flow}
\label{sec:IsoTvsPF}

Now that we have independently verified the simulations of fluid flow (Sec.~\ref{sec:steadyflow}-\ref{sec:vonkarman}) and dendritic growth (Sec.~\ref{sec:isodnn}), we test the coupled model for dendritic growth under convective flow.
Since the current simulations are two-dimensional, quantitative comparisons must be with other 2D calculations, for consistency. 
Hence, we selected quantitative phase-field (PF) results in 2D for equiaxed dendritic growth in a forced flow.

Note that the PF results chosen for comparison, extracted from Refs~\cite{Beckermann99, Tong01}, are for thermal-driven growth at a given undercooling $\Delta$ in a symmetrical model, i.e. with equal solid and liquid diffusivities, while the DNN simulations are for a one-sided model, i.e. neglecting diffusion in the solid phase.
However, normalizing both problem by their respective theoretical steady tip radius and velocity in the diffusive regime, we can directly compare results of the two models.

We simulate the growth of a fourfold equiaxed grain in a forced flow with $\Omega=0.55$ and $\Sch=\nu/D=23.1$, which is equivalent to $\Delta=0.55$ and $\Pra=\nu/D_T=23.1$ in PF calculations of Refs~\cite{Beckermann99, Tong01}, with $D_T$ the thermal diffusivity.
At such a high supersaturation or undercooling, the dendrite tip radius and the diffusion length are of the same order of magnitude --- the Ivantsov Eq.~\eqref{eq:Iv2D} gives $(D/V_s)/R_s\approx2$.
While this makes PF simulations more easily feasible, it does not correspond to the optimal range of application of the DNN method, developed for $R\ll D/V$.
However, as shown in the following, we can still perform reasonable DNN simulations by reducing the grid spacing such that $h\ll D/V$.

In a domain of size $L_x\times L_y=102.4R_s\times70.4R_s$ with $h/R_s=0.1$, we generate an initial nucleus composed of four branches of initial length and tip radius both equal to $R_s$, centered at $x=0.6L_X$ and $y=0.5L_Y$.
We impose an inflow velocity $(u_i,0)$ at the left boundary $(x=0)$, a free outflow with $\partial u/\partial x=\partial v/\partial x=0$ on the right side, symmetry free-slip conditions with $v=0$ and $\partial u/\partial y=0$ along the top and bottom boundaries, as well as mirror symmetry conditions for the $U$ field on all four boundaries.
As in previous simulations, numerical parameters are $K_{\Delta t}=0.5$, $\omega_{\rm up}=0.9$, $\omega_{\rm sor}=1.7$, and $\overline r_{\rm sor}=10^{-3}$.
The thickness of the needles is not bounded ($r_{max}=\infty$), the radius for the integration of the FIF is $r_i=5h$, and branching is set to occur on the side of a needle every time it grows by $10R_s$, i.e. with $L_{sb}=10$ and $\Delta l_{sb}=0$, using notations defined in Sec.~\ref{sec:branching}.
The total simulation time is chosen as $tV_s/R_s=11$, which is sufficient to reach a steady upstream tip velocity. 

\begin{figure}[!b]
\centering
\includegraphics[width=3.in]{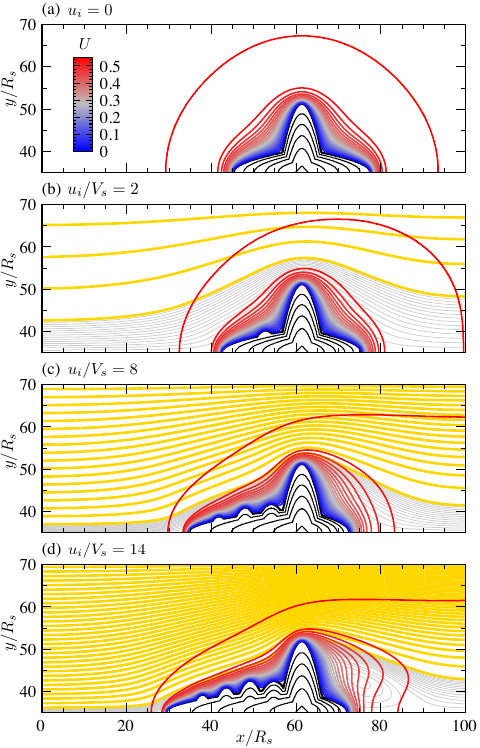}
\caption{
\label{fig:InflowIsoT} 
Results of DNN simulations of an equiaxed grain growth at $\Omega=0.55$, $\Sch=23.1$, and different velocities $u_i$ of the inflow at $x=0$.
For a given $u_i/V_s$, each panel shows: the location of the solid-liquid interface at $tV_s/R_s=0$, 2 ,4, 6, 8, and 10 (black lines); the field $U$ from 0 (blue) to 0.55 (red) with steps of 0.025 at $tV_s/R_s=10$; and streamlines at $tV_s/R_s=10$ (thick yellow and thin gray lines).
Streamlines represent similar iso-values of $|{\rm Str}(x,y)-{\rm Str}_0|$ in all panels, with ${\rm Str}(x,y)$ the integrated stream function (see \ref{app:vortstr}) and ${\rm Str}_0$ the values of ${\rm Str}(x,y)$ in the solid.
Iso-values of $|{\rm Str}(x,y)-{\rm Str}_0|$ are shown from 0 up with steps of 15 (thick yellow lines), as well as in the vicinity of the grain with smaller steps of 1 from 0 to 15 (thin gray lines) in order to highlight the recirculation loops around the downstream branch at high $u_i$.
}
\end{figure}

Fig.~\ref{fig:InflowIsoT} illustrates simulation results at $tV_s/R_s=10$ for inflow velocities $u_i/V_s=0$~(a), 2~(b), 8~(c), and 14~(d).
Thick blue-to-red contour lines show the solute profile.
Streamlines are in thick yellow lines, and in thin gray lines with lower steps in the vicinity of the grain.
Black lines inside the solid region show the time evolution of the grain from $tV_s/R_s=0$ to 10.
Qualitatively, results from DNN simulations exhibit similar features as the corresponding PF calculations, with a grain asymmetry that increases with $u_i$~\cite{Beckermann99, Tong01} and the appearance of recirculation loops in the vicinity of the downstream tip for $u_i/V_s>10$~\cite{Tong01}.

\begin{figure}[!t]
\centering
\includegraphics[width=3.in]{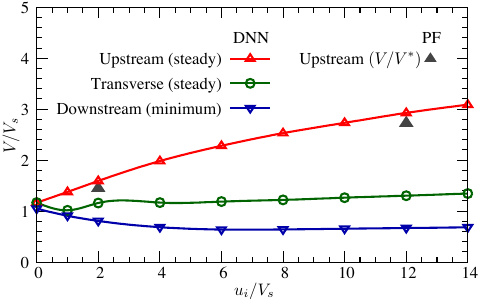}
\caption{
\label{fig:IsoTvsPF} 
Steady velocities of the upstream ($\triangle$) and transverse ($\circ$) tips, and minimum velocity of the downstream tip ($\triangledown$) in DNN simulations, compared to phase-field results of the upstream tip velocity ($\blacktriangle$)~\cite{Tong01} (see values in Table~\ref{tab:scalingPF}).
}
\end{figure}

We make quantitative comparisons of DNN and PF results in Fig.~\ref{fig:IsoTvsPF} in terms of tip growth velocities as a function of the inflow velocity.
In DNN simulations, velocities are averaged over the time range $9\leq tV_s/R_s\leq11$.
The two PF data points correspond to the best converged results (i.e. the lowest interface width $W_0$) from Table~II of Ref.~\cite{Tong01}.
For self-consistency, velocities from PF simulations are scaled with respected to corresponding velocities predicted by exact microscopic solvability theory (here noted $V^*$) as listed in Table~I of Ref.~\cite{Tong01}.
Through this scaling, we can afford to compare PF simulation with symmetric diffusion in solid and liquid phases with the current one-sided approach (liquid diffusion only), and we can also compare results for different strength of the interface anisotropy, namely $\epsilon_{4}=0.05$ and 0.03 in Ref.~\cite{Tong01}.
As summarized in Table~\ref{tab:scalingPF}, the two PF data points correspond to $u_i/V^*\approx2.06$ and $u_i/V^*\approx12.16$, which are comparable to DNN results for $u_i/V_s=2$ and 12.

The discrepancy between PF and DNN results for the upstream tip steady velocity $V_{up}$ is lower than 10\%.
This is satisfactory, considering that these simulations are out of the range of parameters for which the DNN model is expected to remain valid.
In agreement with classical theory and PF results~\cite{Beckermann99, Tong01}, the steady velocity of the transverse tips is weakly affected by the flow (Fig.~\ref{fig:IsoTvsPF}).
Finally, we note that the downstream tip velocity does not reach a steady state.
Instead, the tip velocity reaches a minimum (at a time that decreases with $u_i$) before accelerating due to convective vortices that form around the tip and feed it in solute (see Fig.~\ref{fig:InflowIsoT}d).

\begin{table*}[!t]
\caption{%
Upstream tip steady velocity, $V_{up}$, as a function of the inflow velocity, $u_i$, predicted by DNN simulations compared to PF results~\cite{Tong01}.
DNN results are scaled with respect to the steady state velocity $V_s$ from Eqs~\eqref{eq:R2V} and \eqref{eq:Iv2D}.
PF results are scaled with respect to the exact velocity $V^*$ from microscopic solvability theory, as listed in Ref.~\cite{Tong01}.
}%
\center%
\label{tab:scalingPF}%
\begin{tabular}{ l l l l l l l }%
\hline
\multicolumn{2}{l}{DNN} & \multicolumn{2}{l}{PF} & \multicolumn{3}{l}{From Table~I in Ref.~\cite{Tong01}} \\
$u_i/V_s$ & $V_{up}/V_s$ & $u_i/V^*$ & $V_{up}/V^*$ & $u_id_0/D$ & $V_{up}d_0/D$ & $V^*d_0/D$ \\
\hline
2.0 & {\bf1.59} & 2.06 & {\bf1.45} & 0.035 & 0.0247 & 0.0170 \\ 
12.0 & {\bf2.93} & 12.16 & {\bf 2.73} & 0.135 & 0.0303 & 0.0111 \\ 
\hline
\end{tabular}
\end{table*}

In this specific case, using the DNN method presents little computational advantage compared to PF, since the main advantage of the DNN method comes from the fact that it remains accurate up to $h\approx R_s$, and here $h$ needs to be $\approx0.1R_s$ in order to satisfy $h\ll D/V$.
Thus, each one of these simulations was completed in just under 40~hours with $u_i\neq0$, and less than 12~hours for $u_i=0$, on a single Nvidia Tesla K40c GPU.
Nonetheless, these simulations show that the DNN method still yields reliable predictions of dendritic growth even outside of its most efficient range of parameters, as long as $h\ll D/V$~\cite{dnn3d}.

\subsection{Effect of buoyancy upon the selection of polycrystalline equiaxed microstructure}
\label{sec:Polycrystal}

Finally, we illustrate the potential of the method with simulations of polycrystalline grain growth.
We use physical parameters of an Al-10wt\%~Cu alloy undercooled below its liquidus temperature by $\Delta T=10~$K, which corresponds to a solute supersaturation $\Omega=0.291$ (see \ref{app:parameters} for detailed parameters).  
We perform simulations in the absence of gravity (i.e. pure diffusion), and with buoyancy resulting from Earth gravity (9.81~m/s$^2$).

The simulation domain is a square of $(2.56~$mm$)^2$, initialized at $U=\Omega=0.291$ with 36 solid nuclei.
The nuclei are distributed semi-randomly throughout the domain by dividing it in $6\times6$ squares of equal size, and seeding one grain with random orientation at a random location within each square.

Details of all parameters and their scaling are given in \ref{app:parameters}.
The scaled problem reduces to $\Omega = 0.291$, $\Sch = 252$, $\lambda_c = 0.397$, $\tilde g_x = 0$, $\tilde g_y = 0$ for pure diffusion, and $\tilde g_y = -2294$ for Earth gravity.
The theoretical steady state tip radius is $R_s=2.10~\mu$m and its velocity $V_s=94.8~\mu m$/s.
The numerical grid consists of $608^2$ grid points with a grid spacing $h=2R_s=4.20~\mu$m. 
The simulated time of $10~$seconds corresponds to $451R_s/V_s$.
The half-thickness of the needles is bounded to $r_{max}=2h=4R_s$.
The radius for the integration of the FIF is $r_i=3h$.
Side branching is set to occur every $10\pm2.5R_s$, i.e. with $L_{sb}=10$ and $\Delta l_{sb}=5$.
The initial nuclei are set as small circular seeds composed of four branches of equal length and radius $l_0=R_0=2h$.
External boundary conditions are set to no-flux for solute, and no-slip for fluid velocities.

\begin{figure*}[!t]
\centering
\includegraphics[width=6.6in]{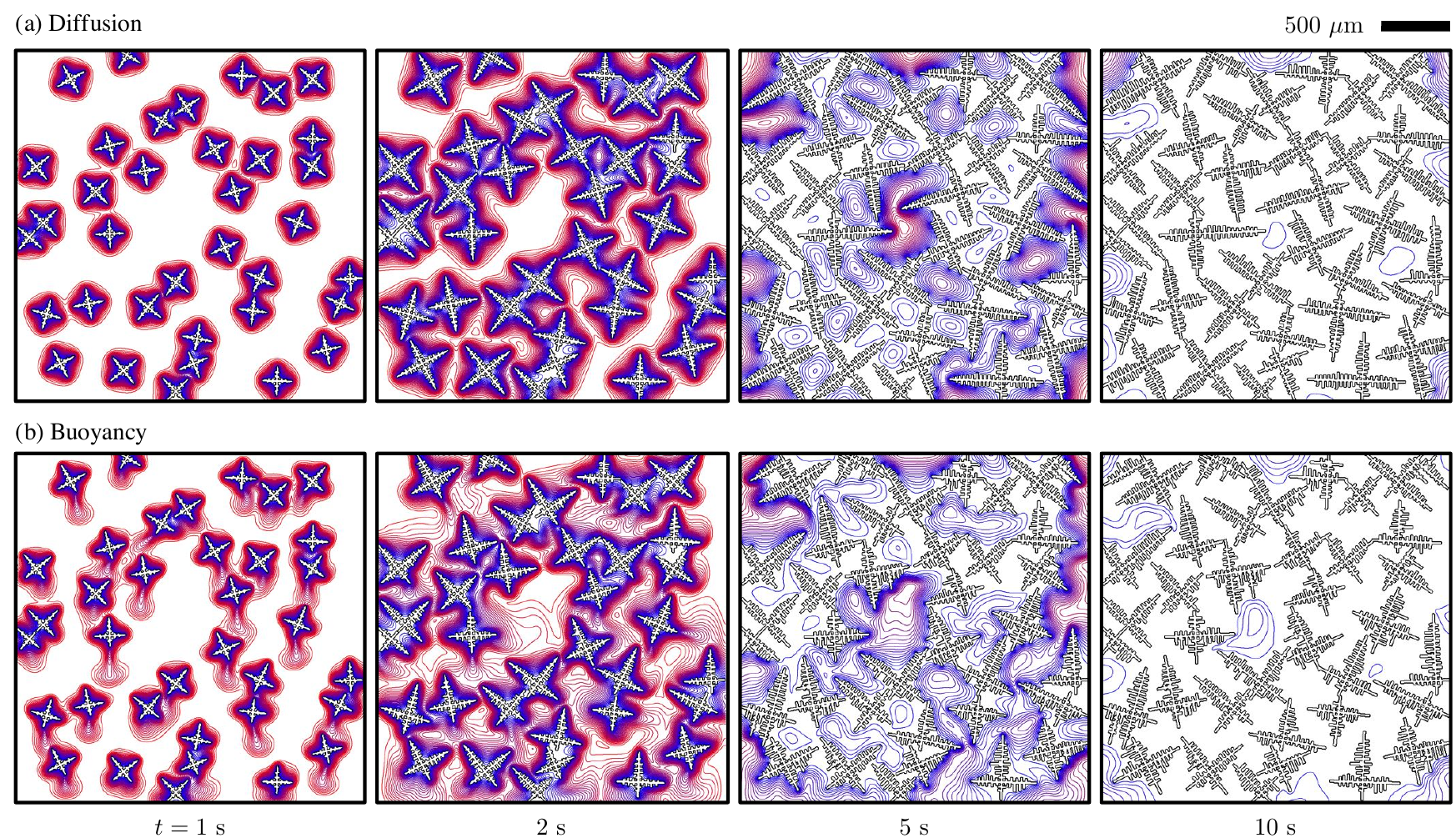}
\caption{
\label{fig:Movies} 
Simulations of polycrystalline growth in Al-10wt\%\,Cu undercooled by 10~K, i.e. with $\Omega=0.291$ in a pure diffusion transport regime (a) and accounting for Earth gravity-induced buoyancy (b).
The solid-liquid interfaces appear as black lines, and color lines represent the solute supersaturation from $U=0$ (blue) to $U=0.29$ (red) with steps of $0.01$.
}
\end{figure*}

The simulations results are shown in Fig.~\ref{fig:Movies} for the simulation in a purely diffusive transport regime (a), and with natural gravity-induced buoyancy (b).
The simulations start with similar initial grain distributions and orientations.
Yet, differences in the solute field appear during crystal growth, which are most noticeable by the presence (or absence) of plumes of sinking solute (Cu)-rich heavy liquid in the early solidification stage (e.g. at $t=1~$s).

\begin{figure*}[!t]
\centering
\includegraphics[width=6.2in]{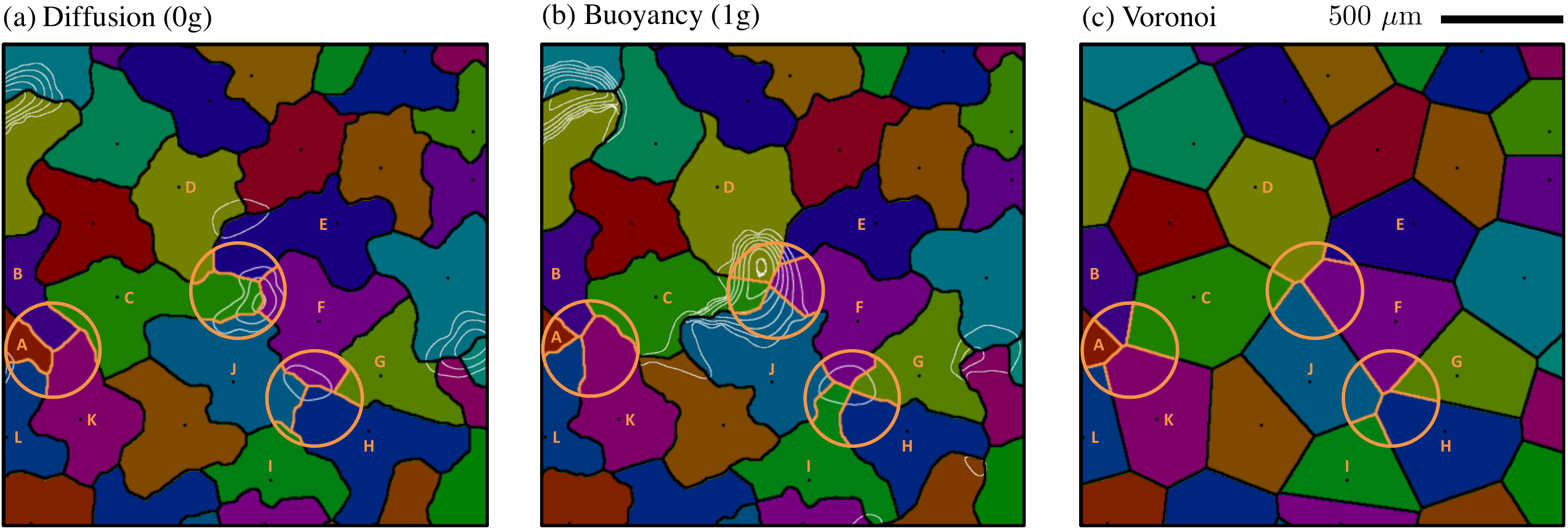}
\caption{
\label{fig:Polycrystal} 
Final grain structures from simulations of Fig.~\ref{fig:Movies}, considering pure diffusion (a) and Earth gravity-driven buoyancy (b), compared to a Voronoi space tessellation (c).
Grain boundaries that differ among simulations are highlighted in orange lines within circled regions.
The locations of initial nuclei are shown as black dots of same size as the nuclei.
Late-solidification highly-segregated regions are shown in white contour lines, corresponding to iso-values of the solute field at $t=10~$s.
The area represented corresponds to the central (2.25~mm)$^2$ region of the simulation domain.
}
\end{figure*}

Even though the size of the domain is limited and the nuclei density is relatively high, the resulting grain structures show some notable differences.
These grain structures are shown in Fig.~\ref{fig:Polycrystal}, and compared to the grain distribution corresponding to a Voronoi space tessellation with similar nuclei locations.
Three regions, marked with orange circles with highlighted grain boundaries, exhibit significant differences in grain structure between solidification under diffusion (a), natural buoyancy (b), or a Voronoi tessellation (c). 
Table~\ref{tab:gb} summarizes which grain boundaries differ between simulations of Figure~\ref{fig:Polycrystal}.
\begin{table}[!b]
\caption{%
Grain boundaries in polycrystalline simulated microstructures: present (\checkmark) or absent (\O) in Figure~\ref{fig:Polycrystal}.
}%
\center%
\label{tab:gb}%
\begin{tabular}{c c c c}%
\hline
Grain & Diffusion & Buoyancy & Voronoi \\
Boundary & (0g) & (1g) &  \\
\hline
A/K & \checkmark & \O & \checkmark \\
B/L & \O & \checkmark & \O \\
\hline
C/E & \checkmark & \O & \O \\
C/F & \checkmark & \checkmark & \O \\
D/F & \O & \checkmark & \checkmark \\
D/J & \O & \O & \checkmark \\
\hline
F/H & \checkmark & \O & \checkmark \\
F/I & \O & \checkmark & \O \\
G/I & \O & \checkmark & \O \\
H/J & \checkmark & \O & \checkmark \\
\hline
\end{tabular}
\end{table}
In this example, it appears that approximating the grain texture by a Voronoi space tessellation (Fig.~\ref{fig:Polycrystal}c) leads to predictions of the existence of grain boundaries that are not in agreement with either diffusive (a) or buoyant (b) simulations.

The different transport regimes also result in different locations of the solute segregated regions, which are represented with white contour lines in Fig.~\ref{fig:Polycrystal}a-b.
Regions of space that are highly segregated in solute are of particular importance in technological metal alloys. 
Solute segregation at grain boundaries change their energy, mobility, and provide preferential sites for the formation of secondary phases that substantially affect the properties of materials~\cite{Raabe14}.
Regions that solidify last are also more prone to solidification defects such as hot tearing.
Here, it is worth noting that the resulting grain structures and segregated regions are only extrapolations of the DNN results, since the approach is not currently capable of accurately representing late stage solidification at high solid fraction. 

While we did not investigate it quantitatively, we have not observed any substantial influence of the initial conditions on the resulting microstructures in terms of nucleus size. However, grain orientations do, as expected, have a critical influence on the resulting microstructures, which is not accounted for when using a Voronoi space tessellation based on an isotropic distance function.  
Furthermore, since all three microstructures in Fig.~\ref{fig:Polycrystal} stem from a unique given nuclei distribution, simulations result in similar grain size distributions. 
However, coupled with an appropriate description of nucleation and thermal conditions (see e.g. Refs~\cite{dnncet_tms, dnncet_jom}), such simulations will allow a deeper investigation into the selection of grain texture and size distribution.

Voronoi space tessellations, such as the one in Figure~\ref{fig:Polycrystal}c, are commonly used as synthetic grain microstructures in polycrystalline micro-mechanical modeling~\cite{Groeber2008,Quey2011}.
They have been used for decades to extract homogenized macroscopic mechanical properties (see e.g. Ref.~\cite{Kumar94}) and more recently used with computational homogenization and crystal plasticity methods (see e.g. Ref.~\cite{Cruzado15}). 
Figure~\ref{fig:Polycrystal} clearly shows that grain structures from dendritic growth modeling are markedly different from their Voronoi counterpart. 
Furthermore, typical Voronoi descriptions of microstructure do not account for the effect of solute segregation at grain boundaries.
Modern micromechanical modeling tools are mature enough to use almost any microstructure morphology as input~\cite{Segurado18}, such that output from DNN simulation could be directly used as input of mechanical modeling of the resulting microstructure.

Simulations were performed on a single Nvidia GTX Titan Xp GPU. 
The simulation in a purely diffusive regime was performed in less than 20~minutes and the one with buoyancy was achieved in under 70~hours. 
The main reason for the difference in computing time stems from the high Schmidt number $\Sch=252$. 
This number illustrates the difference of two orders of magnitude between the diffusivity of solute and that of momentum, i.e. the kinematic viscosity.
The maximum stable time step for an explicit scheme, $\Delta t_{\rm max}$ in Eq.~\eqref{eq:stabdt}, is thus two orders of magnitudes lower with $\Sch=252$ than for purely diffusive conditions, for which $\Delta t_{\rm max}$ is only restricted by the solute diffusivity.
Because a high $\Sch$ makes simulations computationally challenging, simulations of dendritic growth with fluid flow using realistic alloy parameters remain scarce.
For example in Ref.~\cite{Sakane18}, despite using state-of-the-art numerical methods --- namely coupling phase-field and lattice Boltzmann methods and using several hundreds of GPUs --- the alloy viscosity needs to be adjusted --- in this case decreased by a factor 30 --- in order to keep the simulation of a single equiaxed grain in 3D computationally tractable.
Using the DNN method with the relatively straightforward finite difference method described in this article, simulations of a $(2.56~$mm$)^2$ domain with 36 equiaxed grains over 10 seconds and realistic alloy parameters were feasible in just a few days on a single GPU, highlighting their accessibility to any modern desktop computer with a dedicated GPU.
Moreover, like PF simulations~\cite{Sakane18}, these simulations could be accelerated further by using computationally efficient approaches such as highly-parallelizable lattice Boltzmann methods in place of a direct resolution of the Navier-Stokes equations --- however introducing additional parameters (e.g. relaxation time and collision model) to be fitted to actual materials properties. 

In summary, even though present simulations are for the most part illustrative, they clearly demonstrate the potential of the DNN method in quantitatively addressing the influence of fluid flow and gravity-induced buoyancy upon the selection of microstructures during dendritic solidification.

\section{Summary and outlook}
\label{sec:summ}

We presented an extension of the multiscale DNN model that accounts for convective transport in the liquid phase. 
We proposed a first formulation of the model for isothermal growth of a binary alloy in a solute-supersaturated liquid.
The numerical implementation is based on standard numerical methods, namely combining: (i) finite differences on a regular square grid, (ii) a mostly explicit time stepping scheme, only with the incompressibility condition treated iteratively, and (iii) an approximate step-wise geometrical description of parabolic dendritic branches.
We introduced the use of a circular contour for the integration of the flux intensity factor at the tip of each needle.
The latter seamlessly allows the simulation of dendritic branches of arbitrary orientation independently from the numerical grid, which enables, for instance, the simulation of hexagonal crystal structures and that of polycrystalline microstructures.

We verified the predictions of the model against established test cases independently for fluid flow (steady, buoyant, and unsteady with an obstacle) and for crystal growth (steady-state for various solute supersaturations and growth directions).
We performed simulations of equiaxed growth of a single grain in a forced flow, comparing with published quantitative phase-field results.
In spite of the relatively straightforward numerical methods employed, the current code yields predictions within 10\% of phase-field predictions for tip growth velocities.
Finally, we performed an illustrative simulation of polycrystalline grain growth using physical alloy parameters for a Al-10wt\%Cu alloy.
Resulting grain structures show notable differences, depending upon whether gravity-driven buoyancy is taken into account or not.

The current model opens the way to a number of theoretical studies on the effect of fluid flow on dendritic growth at experimentally-relevant length and time scales.
These studies include investigating the effect of flow conditions and relative orientation on the steady and transient growth of an equiaxed grain~\cite{CantorVogel77, Ananth91, Sakane18, Badillo07}.
Further quantitative comparisons of synthetic microstructures from solidification modeling and Voronoi-based representations, such as the one illustrated in Section~\ref{sec:Polycrystal}, would be useful to clearly identify cases where simple geometrical representations of grain structures are sufficient.
In the framework of building integrated virtual processing and testing tools, further efforts in coupling micromechanical simulations using input from solidification models such as DNN are also needed. 

Some extensions of the current model, without any conceptual change to the model, would be particularly useful. 
The most important is the extension to three dimensions, which is essential, since the flow pattern and resulting crystal growth dynamics are fundamentally different in 3D and in 2D~\cite{Jeong01, Jeong03, YuanLee10}. 
This difference was already shown to be crucial for quantitative predictions in the purely diffusive regime~\cite{mcwasp}.
Because of that difference, results of a model in 2D cannot be quantitatively compared to experimental data, which is by essence three-dimensional.

A second direct extension of the model is the application to directional solidification conditions. 
It could be done without any major conceptual change to the model in the same manner as it was done for pure diffusion, i.e. using a frozen temperature approximation~\cite{dnn2d,dnn3d}.
Such an extension would allow for mapping of entire ranges of crystalline orientations, as well as amplitudes and relative orientations of the temperature gradient and gravity.
One could thus investigate the influence of these processing conditions upon the selection of inner grain spacings~\cite{Steinbach09} and that of orientation and roughness of grain boundaries~\cite{Tourret15, Tourret17, Pineau18}.

Further extensions directly relate to limitations of the current DNN formulation.
New formulations are required to ensure proper mass and solute conservation at all times.
This important requirement is not currently fulfilled, due to the combined effects of (i) the assumption of a Laplacian field within the integration domain around each tip (Eq.~\eqref{eq:Laplace}), (ii) the bounding of each needle thickness far away from the tip, and (iii) the fact that sidebranches are added phenomenologically.
Mass conservation is key for mostly two important reasons, namely to: (i) provide a better description of late-stage solidification when the solid fraction approaches unity, and (ii) appropriately estimate the mass of solid grains in order to calculate their buoyant motion in the liquid by floatation or sinking.
Mass conservation in the DNN approach was not addressed in the current article, as it is the focus of ongoing work and will be addressed elsewhere.

Additionally, quantitative benchmarks to other multiscale modeling approaches for dendritic growth would also be particularly useful in order to clarify the strengths, weaknesses, and to guide the selection of appropriate numerical parameters for each approach. Such studies are currently underway.

In summary, we demonstrated that the dendritic needle network approach could be readily applied to convective transport in the liquid phase.
The DNN model yields predictions in good agreement with the more computationally demanding phase-field method.
The next two steps for the method are its extensions to (i) directional solidification conditions and (ii) three dimensions. 
These two extensions, which do not require any conceptual change to the model, have already been performed for purely diffusive transport, and are currently underway.
The DNN model provides a simple, efficient, and quantitative way of investigating dendritic growth at an intermediate scale between phase-field and coarse-grained volume-averaged models.
Further utilization and extension of the approach will certainly shed light upon open questions on the influence of convection on dendritic microstructure selection, e.g. selection of spacings and of grain boundaries, for realistic physical parameters and at length and time scales directly comparable to experiments.

\section*{Acknowledgements}
This work was supported by a Postdoctoral Director's Fellowship from the U.S. Department of Energy through Los Alamos National Laboratory LDRD Program.
A.J.C. acknowledges support from the U.S. DOE Office of Science, Office of Basic Energy Sciences, Materials Sciences and Engineering Division, under Contract Number DE-SC001606. 
For the writing of the article, D.T. was supported by the Advanced Grant VIRMETAL of the European Research Council under the European Union's Horizon 2020 research and innovation programme (Grant agreement number 669141).
D.T. also gratefully acknowledges support by the NVIDIA Corporation with the donation of a Titan Xp GPU.

\begin{appendix} 

\section{Notations}
\label{app:symbols}

\newpage

\setcounter{figure}{0}
\setcounter{table}{0}

\begin{table*}[!t]
\caption{%
Notations used in the article --- Latin symbols.
}%
\center%
\label{tab:init}%
\begin{tabular}{| l l c |}%
\hline
Symbol & Definition  & Value [Reference] \\
\hline
$a$ & Integration distance from the tip & [Figs~\ref{fig:DNN}, \ref{fig:fif}]\\
$\tilde a$ & Integration distance from the tip | scaled & $a/R_s$ \\
$c$ & Solute concentration field & \\
$c_0$ & Liquid equilibrium concentration at the reference temperature $T_0$& [Fig.~\ref{fig:appd1}] \\
$c_\infty$ & Nominal solute concentration of the alloy & [Fig.~\ref{fig:appd1}] \\
$d$ & Obstacle diameter & [Sec.~\ref{sec:vonkarman}] \\
$d_0$ & Solute capillary length of the solid-liquid interface at $T_0$ & [Eq.~\eqref{eq:capillength}] \\
$f$ & Outflow oscillation frequency & [Sec.~\ref{sec:vonkarman}] \\
$f_\gamma$ & Interface stiffness anisotropy function & [Sec.~\ref{sec:sharpintpb}]\\
$f_\mu$ & Body forces component in the $\mu\in \{x,y\}$ direction & [Eq.~\eqref{eq:bodyf}] \\
$\gr$ & Gravity vector field & \\
$\tilde\gr$ & Gravity vector field $(\tilde g_x,\tilde g_y)$ | scaled & $\gr\,R_s/V_s^2$\\
$\tilde g_\mu$ & Gravity component in the $\mu\in \{x,y\}$ direction | scaled & \\
$h$ & Finite difference grid element size & \\
$k$ & Interface solute partition coefficient & [Fig.~\ref{fig:appd1}] \\
$l_D$ & Solute diffusion length in the liquid & $D/V$ \\
$l_{sb}$ & Distance to grow until the next sidebranching event & $\big(L_{sb}+\delta l_{sb}\big)R_s$\\
$m$ & Alloy liquidus slope | here with $m<0$& [Fig.~\ref{fig:appd1}] \\
$n$ & Unit normal vector component & \\
${\mathbf n}$ & Unit normal vector & \\
$p$ & Pressure field & \\
$q_{i,j}$ & Tip flux integration surface mask function & $(i,j)\in\Sigma_i$ [Sec.~\ref{sec:growth}]\\
$r_{max}$ & Maximal (bounded) half-width of each needle & [Sec.~\ref{sec:growth}] \\	
$r_{\rm sor}$ & SOR current iteration residual & [Eq.~\eqref{eq:residual}]\\
$\overline r_{\rm sor}$ & SOR residual required for convergence & [Sec.~\ref{sec:parallel}] \\
$t$ & Time & \\	
$\uu$ & Fluid velocity vector field & \\	
$u$ & Fluid velocity component along the $x$ direction & \\
$u_i$ & Imposed fluid velocity at a boundary & [Sec.~\ref{sec:vonkarman}, \ref{sec:IsoTvsPF}]\\	
$v$ & Fluid velocity component along the $y$ direction & \\		
$\vv$ & Fluid velocity vector field | scaled & $\uu/V_s$\\
$v_n$ & Solid-liquid interface normal velocity & [Eq.~\eqref{eq:StefanU}]\\
$x$ & Cartesian space coordinate & \\	
$x_t$ & Tip location along the $x$ direction & [Sec.~\ref{sec:intermedscale}]\\
$y$ & Cartesian space coordinate & \\
$y_i$ & Interface location along the $y$ direction & [Sec.~\ref{sec:intermedscale}] \\		
\hline
\end{tabular}
\\{\it [Table continued on next page]}
\end{table*}
\begin{table*}[!t]
\center%
\label{tab:init}%
\begin{tabular}{| l l c |}%
\hline
Symbol & Definition  & Value [Reference] \\
\hline
$D$ & Liquid solute diffusivity & \\
$D_T$ & Thermal diffusivity & \\
${\mathcal F}$  & Tip flux intensity factor at a dendrite tip & [Eq.~\eqref{eq:newFIF2D}]\\
$\tilde{\mathcal F}$ & Scaled flux intensity factor & [Eq.~\eqref{eq:FIF_scaled}]\\
${\bf F_V}$ & Body forces vector field & [Eq.~\eqref{eq:momentum}]\\
${\rm Iv}$ & Ivantsov function & [Eq.~\eqref{eq:Iv2D}] \\
$K_{\Delta t}$ & User-input safety factor on the adaptive time step selection & [Eq.~\eqref{eq:adaptdt}]\\
$L_{sb}$ & Average distance between consecutive sidebranches & [Sec.~\ref{sec:branching}]\\
$L_x$ & Simulation domain size in the $x$ direction & \\
$L_y$ & Simulation domain size in the $y$ direction & \\
$N$ & Number of inner grid points in a given direction & [Sec.~\ref{sec:bc}]\\
$N_{\rm iter}$ & Maximum number of SOR iterations to solve Eq.~\eqref{eq:poisson}& [Sec.~\ref{sec:parallel}]\\
$R$ & Dendrite tip radius & [Fig.~\ref{fig:DNN}]\\
$\tilde R$ & Dendrite tip radius | scaled & $R/R_s$ \\
$R_s$ & Steady state free dendrite tip radius & [Sec.~\ref{sec:scaling}] \\
$R_g$ & Ideal gas constant & 8.314 J\,K$^{-1}$mol$^{-1}$ \\
${\rm Str}$ & Flow stream function & [Eq.~\eqref{eq:stream}] \\
${\rm Str}_0$ & Value of the stream function in the solid & [Fig.~\ref{fig:InflowIsoT}] \\	
$T$ & Temperature & \\	
$T_L$ & Liquidus temperature of the alloy & [Fig.~\ref{fig:appd1}] \\		
$T_M$ & Melting temperature of the pure solvent & [Fig.~\ref{fig:appd1}] \\			
$T_S$ & Solidus temperature of the alloy & [Fig.~\ref{fig:appd1}] \\			
$T_0$ & Temperature of the domain & [Fig.~\ref{fig:appd1}] \\	
$U$ & Solute concentration field | scaled & [Eq.~\eqref{eq:defU}]\\
$U_i$ & Equilibrium solute concentration at the liquid-solid interface & [Eqs~\eqref{eq:GT},\eqref{eq:Uequal0}]\\	
$V$ & Dendrite tip growth velocity & [Fig.~\ref{fig:DNN}] \\	
$\tilde V$ & Dendrite tip growth velocity | scaled & $V/V_s$ \\	
$V_s$ & Steady state free dendrite tip velocity & [Sec.~\ref{sec:scaling}] \\	
$V^*$ & Dendrite tip velocity from microscopic solvability theory & [Sec.~\ref{sec:IsoTvsPF}] \\
$V_{up}$ & Upstream dendrite tip steady velocity & [Sec.~\ref{sec:IsoTvsPF}] \\
${\rm Vort}$ & Flow vorticity field & [Eq.~\eqref{eq:vort}] \\
\hline
\end{tabular}
\end{table*}

\clearpage

\begin{table*}[!t]
\caption{%
Notations used in the article --- Greek symbols.
}%
\center%
\label{tab:init}%
\begin{tabular}{| l l c |}%
\hline
Symbol & Definition & Value [Reference] \\
\hline
$\alpha$ & Liquid solute diffusivity | scaled & $D/(R_s V_s)$ \\
$\beta_c$ & Liquid solutal expansion coefficient & [Eq.~\eqref{eq:expcoeff}] \\ 				
$\gamma$ & Orientation-dependent excess free energy of the solid-liquid interface & [Sec.~\ref{sec:sharpintpb}] \\ 
$\gamma_0$ & Averaged value of $\gamma(\bar\theta)$ over all orientations in a (100) plane & [Sec.~\ref{sec:sharpintpb}] \\ 
$\delta l_{sb}$ & Distance fluctuation between consecutive sidebranches & $\in [-\Delta l_{sb}/2;\Delta l_{sb}/2]$\\ 			
$\epsilon_4$ & Strength of solid-liquid interface excess free energy anisotropy & \\ 			
$\eta$ & Dynamic fluid viscosity & \\
$\theta$ & Dendrite growth orientation angle with the $x$ direction & [Fig.~\ref{fig:fif}] \\
$\bar\theta$ & Solid-liquid interface orientation & \\
$\kappa$ & Curvature of the solid-liquid interface & \\ 				
$\lambda_c$ & Liquid solutal expansion coefficient | scaled & $(1-k)c_0\beta_c$ \\
$\nu$ & Liquid kinematic viscosity & $\eta/\varrho_0$ \\
$\xi$ & Field (any) & \\
$\varrho$ & Fluid density & \\
$\varrho_0$ & Fluid density in a reference state with $T=T_0$  and $c=c_0$ & \\
$\sigma$ & Dendrite tip selection parameter & [Eq.~\eqref{eq:R2V}] \\
$\tau$ & Time | scaled & $t V_s/R_s$ \\
$\chi$ & Liquid kinematic viscosity | scaled & $\nu/(R_s V_s)$\\
$\psi$ & Pressure field | scaled & $p/(\varrho_0 V_s^2)$ \\
$\omega_{\rm sor}$ & Successive over relaxation (SOR) relaxation parameter & [Sec.~\ref{sec:timestepping}] \\
$\omega_{\rm up}$ & Upwind parameter for the spatial discretization of convective terms & [Sec.~\ref{sec:spacediscr}] \\
&&\\%
$\Gamma_0$ & Solute flux integration contour along the solid-liquid interface & [Figs~\ref{fig:DNN}, \ref{fig:fif}] \\ 			
$\Gamma_i$ & Solute flux integration contour around the tip & [Figs~\ref{fig:DNN}, \ref{fig:fif}] \\ 			
$\Gamma_{sl}$ & Gibbs-Thomson coefficient of the solid-liquid interface & \\ 			
$\Delta$ & Dimensionless liquid undercooling & [Eq.~\eqref{eq:delta}]\\ 				
$\Delta l_{sb}$ & Amplitude of the distance fluctuation between consecutive sidebranches & \\ 			
$\Delta t$ & Numerical time step & \\ 				
$\Delta T$ & Alloy liquid undercooling & $T_L-T_0$ \\ 				
$\Delta T_0$ & Alloy unit undercooling & $T_L-T_S$ \\ 				
$\Delta t_{\rm max}$ & Maximum stable time step for the explicit Euler scheme & [Eq.~\eqref{eq:stabdt}]\\ 		
$\Sigma_i$ & Solute flux integration surface around the tip & [Figs~\ref{fig:DNN}, \ref{fig:fif}] \\
$\Omega$ & Dimensionless liquid solute supersaturation & [Eq.~\eqref{eq:omega}]\\
\hline
\end{tabular}
\end{table*}

\begin{table*}[!t]
\caption{%
Notations used in the article --- superscripts and subscripts.
}%
\center%
\label{tab:init}%
\begin{tabular}{| l l c |}%
\hline
Symbol & Definition & Value [Reference] \\
\hline
$bc$ & Imposed boundary conditions & [Sec.~\ref{sec:bc}]\\ 		
$c$ & Centered difference spatial discretization & [\ref{app:discrsolute}] \\ 		
$dc$ & Donor-cell scheme spatial discretization & [\ref{app:discrsolute}] \\ 		
$i$ & Integer finite difference grid index along the $x$ direction & \\ 		
$\ii$ & Staggered grid index shifted in the $x$ direction & $i+1/2$ [Fig.~\ref{fig:grid}] \\
${(it)}$ & Current SOR iteration & [\ref{app:discrpoisson}] \\ 	
${(it+1)}$ & Next SOR iteration & [\ref{app:discrpoisson}] \\
$j$  & Integer finite difference grid index along the $y$ direction & \\
$\jj$ & Staggered grid index shifted in the $y$ direction & $j+1/2$ [Fig.~\ref{fig:grid}] \\
${(n)}$ & Current time step & [Sec.~\ref{sec:timestepping}] \\
${(n+1)}$ & Next time step & [Sec.~\ref{sec:timestepping}] \\
${(*)}$ & Intermediate step velocities solution of Eq.~\eqref{eq:projector} & [Sec.~\ref{sec:timestepping}] \\
\hline
\end{tabular}
\end{table*}

\begin{table*}[!t]
\caption{%
Notations used in the article --- classical nondimensional numbers.
}%
\center%
\label{tab:init}%
\begin{tabular}{| l l c |}%
\hline
Symbol & Definition & Value [Reference] \\
\hline
$\Nus$ & Nusselt number  & $\Big(\int_0^{L} \partial_x U ~{\rm d}y\Big)/L$ [Sec.~\ref{sec:steadybuoyancy}]\\	
$\Pec$ & P\'eclet number  & $R V/(2D)$ \\		
$\Pra$ & Prandtl number & $\nu/D_T$ \\	
$\Ray$ & Rayleigh number & $\beta_c|\gr|/(D\nu)$ \\
$\Rey$ & Reynolds number relative to a characteristic fluid velocity $u_0$ & $R_s u_0/\nu$\\	
$\Rey^*$ & Reynolds number relative to the crystal growth velocity $V_s$ & $ R_sV_s/\nu$ \\	
$\Sch$ & Schmidt number & $\nu/D$ \\	
$\Str$ & Strouhal number & $f d/ u_i$ [Sec.~\ref{sec:vonkarman}] \\	
\hline
\end{tabular}
\end{table*}

\clearpage 

\section{Discretized equations}
\setcounter{figure}{0}
\setcounter{table}{0}
\label{app:discr}

\subsection{Solute conservation}
\label{app:discrsolute}

Diffusive terms in Eq.~\eqref{eq:U2D} are discretized using centered finite differences as
\begin{align}
\left[ \nabla^2 U \right]_{i,j} \approx \frac{U_{i+1,j} + U_{i-1,j} + U_{i,j+1} + U_{i,j-1} - 4\,U_{i,j}}{h^2} .
\end{align}
For convective terms we use a weighted average of centered difference and donor-cell scheme, e.g. for the term along the $x$ direction
\begin{align}
\label{eq:weightedx}
 \partial_x (u\,U) &= \omega_{\rm up} \Big[ \partial_x (u\,U) \Big]^{dc} + (1-\omega_{\rm up}) \Big[ \partial_x (u\,U) \Big]^{c} ,
\end{align}
where the centered difference term
\begin{align}
\Big[ \partial_x (u\,U) \Big]^{c}_{i,j} = \frac{1}{h}\Big[ u_{i+\frac{1}{2},j}\tilde U_{i+\frac{1}{2},j} - u_{i-\frac{1}{2},j}\tilde U_{i-\frac{1}{2},j}\Big], 
\end{align}
with solute values centered between grid points
\begin{align}
\tilde U_{i+\frac{1}{2},j} = \left( U_{i,j} +U_{i+1,j} \right)/2 ,\\
\tilde U_{i-\frac{1}{2},j} = \left( U_{i,j} +U_{i-1,j} \right)/2 ,
\end{align}
and the donor-cell term may be written~\cite{gdn}
\begin{align}
\Big[ \partial_x (u\,U) \Big]^{dc}_{i,j} = \frac{1}{2h}\Big[ & u_{i+\frac{1}{2},j}\big( U_{i,j} +U_{i+1,j} \big) \nonumber\\&
+ \left| u_{i+\frac{1}{2},j}\right| \big( U_{i,j} - U_{i+1,j}\big) \nonumber\\&
- u_{i-\frac{1}{2},j}\big( U_{i,j} +U_{i-1,j}\big) \nonumber\\&
+ \left| u_{i-\frac{1}{2},j}\right| \big( U_{i,j} - U_{i-1,j}\big) \Big].
\end{align}
Developing Eq.~\eqref{eq:weightedx}, simplifying, and introducing the shifted grid index $\ii\equiv i+1/2$ for the $u$ field (see Fig.~\ref{fig:grid}a), this yields
\begin{align}
\Big[ \partial_x (u\,U) \Big]_{i,j} &= \Big[ \partial_x (u\,U) \Big]^{c}_{i,j} \nonumber\\&
+\frac{\omega_{\rm up}}{2h} \bigg[ \left| u_{\ii,j}\right| \big( U_{i,j} - U_{i+1,j}\big)\nonumber\\&
+ \left| u_{\ii-1,j}\right| \big( U_{i,j} - U_{i-1,j}\big)\bigg].
\end{align}
Similarly, along the $y$ direction
\begin{align}
\Big[ \partial_y (v\,U) \Big]_{i,j} &= \Big[ \partial_y (v\,U) \Big]^{c}_{i,j} \nonumber\\&
+\frac{\omega_{\rm up}}{2h} \bigg[ \left| v_{i,\jj}\right| \big( U_{i,j} - U_{i,j+1}\big)\nonumber\\&
+ \left| v_{i,\jj-1}\right| \big( U_{i,j} - U_{i,j-1}\big)\bigg].
\end{align}
where $\jj\equiv j+1/2$ is the shifted grid index for the $v$ field (Fig.~\ref{fig:grid}a).

\subsection{Momentum conservation}
\label{app:discrpredict}

Like in \ref{app:discrsolute} and Ref.~\cite{gdn}, we use centered differences for diffusive terms and weighted average of centered differences and donor-cell scheme for convective terms.
Discretized terms of Eq.~\eqref{eq:u2D} thus write
\begin{align}
\Big[ \partial_{xx} u \Big]_{\ii,j} =~& \frac{1}{h^2} \big( u_{\ii+1,j} +u_{\ii-1,j} -2 u_{\ii,j} \big) \\
\Big[ \partial_{yy} u \Big]_{\ii,j} =~& \frac{1}{h^2} \big( u_{\ii,j+1} +u_{\ii,j-1} -2 u_{\ii,j} \big) 
\end{align}
\begin{align}
\Big[ \partial_x (u^2) \Big]_{\ii,j} =~& \frac{1}{4h} \Big[ \big( u_{\ii,j} +u_{\ii+1,j} \big)^2 -\big( u_{\ii,j} +u_{\ii-1,j} \big)^2  \nonumber\\ &
+ \omega_{\rm up} \left| u_{\ii,j} +u_{\ii+1,j} \right| \big ( u_{\ii,j} - u_{\ii+1,j} \big)\nonumber\\ &
+ \omega_{\rm up} \left| u_{\ii,j} +u_{\ii-1,j} \right| \big ( u_{\ii,j} - u_{\ii-1,j} \big)  \Big] 
\end{align}
\begin{align}
\Big[ \partial_y (uv) \Big]_{\ii,j} =~& \frac{1}{4h} \Big[ 
  \big( u_{\ii,j} +u_{\ii,j+1} \big) \big(v_{i,\jj}+v_{i+1,\jj}\big) \nonumber\\ &
- \big( u_{\ii,j} +u_{\ii,j-1} \big) \big(v_{i,\jj-1}+v_{i+1,\jj-1}\big) \nonumber\\ &
+ \omega_{\rm up} \left| v_{i, \jj} +v_{i+1, \jj} \right| \big ( u_{\ii,j} - u_{\ii,j+1} \big) \nonumber\\ &
+ \omega_{\rm up} \left| v_{i, \jj-1} +v_{i+1, \jj-1} \right| \big ( u_{\ii,j} - u_{\ii,j-1} \big)  \Big] 
\end{align}
\begin{align}
\Big[ f_x(U) \Big]_{\ii,j} =~&  \big( 1 + \lambda_c U_{\ii,j} \big) \tilde g_x
\end{align}
where $\ii\equiv i+1/2$ and $\jj\equiv j+1/2$ are shifted grid indices for the $u$ and $v$ fields, respectively (Fig.~\ref{fig:grid}a), and hence $U_{\ii,j}=(U_{i,j}+U_{i+1,j})/2$.

Similarly terms of Eq.~\eqref{eq:v2D} are discretized
\begin{align}
\Big[ \partial_{xx} v \Big]_{i,\jj} =~& \frac{1}{h^2} \big( v_{i+1,\jj} +v_{i-1,\jj} -2 v_{i,\jj} \big) \\
\Big[ \partial_{yy} v \Big]_{i,\jj} =~& \frac{1}{h^2} \big( v_{i,\jj+1} +v_{i,\jj-1} -2 v_{i,\jj} \big) 
\end{align}
\begin{align}
\Big[ \partial_y (v^2) \Big]_{i,\jj} =~& \frac{1}{4h} \Big[ \big( v_{i,\jj} + v_{i,\jj+1} \big)^2 - \big( v_{i,\jj} +v_{i,\jj-1} \big)^2  \nonumber\\ &
+ \omega_{\rm up} \left| v_{i,\jj} +v_{i,\jj+1} \right| \big ( v_{i,\jj} - v_{i,\jj+1} \big)\nonumber\\ &
+ \omega_{\rm up} \left| v_{i,\jj} +v_{i,\jj-1} \right| \big ( v_{i,\jj} - v_{i,\jj-1} \big)  \Big] 
\end{align}
\begin{align}
\Big[ \partial_x (uv) \Big]_{i,\jj} =~& \frac{1}{4h} \Big[ 
  \big( u_{\ii,j} +u_{\ii,j+1} \big) \big(v_{i,\jj}+v_{i+1,\jj}\big) \nonumber\\ &
- \big( u_{\ii-1,j} +u_{\ii-1,j+1} \big) \big(v_{i,\jj}+v_{i-1,\jj}\big) \nonumber\\ &
+ \omega_{\rm up} \left| u_{\ii,j} + u_{\ii,j+1} \right| \big ( v_{i, \jj} - v_{i+1, \jj} \big) \nonumber\\ &
+ \omega_{\rm up} \left| u_{\ii-1,j} + u_{\ii-1,j+1} \right| \big ( v_{i, \jj} - v_{i-1, \jj} \big)  \Big] 
\end{align}
\begin{align}
\Big[ f_y(U) \Big]_{i,\jj} =~& \big( 1 + \lambda_c U_{i,\jj} \big)  \tilde g_y.
\end{align}

\subsection{Pressure Poisson}
\label{app:discrpoisson}

We solve Eq.\eqref{eq:poisson} by iteratively applying
\begin{align}
\label{eq:iterSOR}
\psi^{(it+1)}_{i,j} &= (1-\omega_{\rm sor})\,\psi_{i,j}^{(it)} \nonumber \\&
+ \frac{\omega_{\rm sor}}{4}\bigg\{ \big[ \psi_{i+1,j}+\psi_{i-1,j}+\psi_{i,j+1}+\psi_{i,j-1}\big]^{(it)} \nonumber\\&
- \frac{h}{\Delta t}\big[ u_{\ii,j} - u_{\ii-1,j} + v_{i,\jj} - v_{i,\jj-1} \big]^{(*)} \bigg\}
\end{align}
where $\ii\equiv i+1/2$, $\jj\equiv j+1/2$, superscripts $^{(it)}$ and $^{(it+1)}$ denote successive iteration steps, and $^{(*)}$ denotes velocity components solution of the pressure-less momentum conservation equation from the previous step in the temporal loop (Step~3 in Table~\ref{tab:algo}).

The residual $r_{\rm sor}$, defined as Eq.~\eqref{eq:residual}, is calculated using CUDA's {\it atomicAdd} operator in order to avoid race conditions among parallel threads.
The denominator ${|| \psi^{(n)} ||_2}$ is also calculated using {\it atomicAdd} during step~(6) of Table~\ref{tab:algo}, i.e. when updating velocities $u^{(n+1)}$, $v^{(n+1)}$ at the end of the previous time step.

\section{Circular contour intersection}
\setcounter{figure}{0}
\setcounter{table}{0}
\label{app:acirc}

\begin{figure}[!h]
\centering
\includegraphics[width=2in]{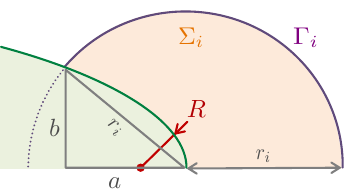}
\caption{
\label{fig:appb1} 
Intersection of a circular integration contour $\Gamma_i$ of radius $r_i$ and a parabolic needle tip of radius $R$.
}
\end{figure}

On the schematics in Fig.~\ref{fig:appb1}, the gray right triangle has $r_i^2=a^2+b^2$.
Yet, the half-width of the parabola at a distance $a$ from its tip is $b=\sqrt{2R a}$, which yields $a^2+2R a-r_i^2 =0$.
Solving the latter for $a\geq0$ yields
\begin{align}
a=-R+\sqrt{R^2+r_i^2}.
\end{align}

%

\section{Vorticity and stream function}
\setcounter{figure}{0}
\setcounter{table}{0}
\label{app:vortstr}

Vorticity and stream function are here only used for visualization purposes and not for calculation of the fluid dynamics.
We use them for instance to render streamlines in Fig.~\ref{fig:ghia} and \ref{fig:InflowIsoT} as iso-values of the stream function.
Vorticity field and stream function are respectively defined as
\begin{equation}
\label{eq:vort}
{\rm Vort}(x,y) \equiv \frac{\partial u}{\partial y} - \frac{\partial v}{\partial x} , 
\end{equation}
\begin{equation}
\label{eq:stream}
\frac{\partial {\rm Str}(x,y)}{\partial x} \equiv -v, \quad \frac{\partial {\rm Str}(x,y)}{\partial y} \equiv u.
\end{equation}
Numerically, the vorticity can be calculated directly from the local velocity field, and the stream function can be integrated, for instance setting ${\rm Str}_{i,0}=0$ and integrating over $j$.
For convenience, we calculate those fields at the center of finite difference elements (i.e. at the corner of the equivalent finite volume cells in Fig.~\ref{fig:grid}a)~\cite{gdn}.

\section{Polycrystalline simulation parameters}
\setcounter{figure}{0}
\setcounter{table}{0}
\label{app:parameters}

The supersaturation and buoyancy conditions in the simulation of Section~\ref{sec:Polycrystal} are completely defined by 
$\Omega = 0.291$,
$\Sch = 252$,
$\lambda_c = 0.397$,
$g_x = 0$, and
$g_y = -2294$ for Earth level gravity, or $g_y = 0$ for pure diffusion. 
In the following subsections, we derive those scaled parameters and provide details on numerical parameters and initial conditions.

\begin{table*}[!t]
\caption{%
Alloy parameters for Al-Cu ($T$ in K in formulas).}%
\center%
\label{tab:alcu}%
\begin{tabular}{l l r l c}%
\hline
Property & Symbol & Value & Unit & Ref.\\
\hline
Composition (Cu) & $c_\infty$ & 10 & wt\% & \\
Undercooling & $\Delta T$ & 10 & K & \\
Melting point (Al) & $T_M$ & 933 & K & \\
Solute partition coefficient & $k$ & $0.14$ &  & \cite{KurzFisher92}\\
Liquidus slope & $m$ & $-3.0$ & K\,wt\%$^{-1}$ & \cite{clarke2017}\\
Liquid solute diffusivity & $D$ & $ 2.4\times 10^{-9}$ & m$^2$\,s$^{-1}$ & \cite{LeeLiuMiyaharaTrivedi04}\\
Gibbs-Thomson coefficient & $\Gamma_{sl}$ & $2.4\times10^{-7}$ & Km & \cite{KurzFisher92,GunduzHunt85}\\
Interface anisotropy strength & $\epsilon_{4}$ & 0.012 &  & \cite{Morris02}\\
Liquid density & $\varrho(Al-4wt\%Cu)$ & $2.43-3.2\times 10^{-4}(T-922)$ & g\,cm$^{-3}$ & \cite{Plevachuk08}\\
 & $\varrho(Al-20wt\%Cu)$ & $2.71-4.05\times 10^{-4}(T-873)$ & g\,cm$^{-3}$ & \cite{Plevachuk08}\\
Liquid viscosity & $\eta$(Al-4wt\%Cu) & 0.196\,$\exp$\{15206/$(R_gT)$\} & mPa\,s & \cite{Plevachuk08}\\
 & $\eta$(Al-20wt\%Cu) & 0.209\,$\exp$\{15295/$(R_gT)$\} & mPa\,s & \cite{Plevachuk08}\\
Gas constant & $R_g$ & 8.314 & J\,K$^{-1}$mol$^{-1}$ & \\
\hline
\end{tabular}
\end{table*}
%

\subsection{Undercooling and supersaturation}

\begin{figure}[!b]
\centering
\includegraphics[width=1.8in]{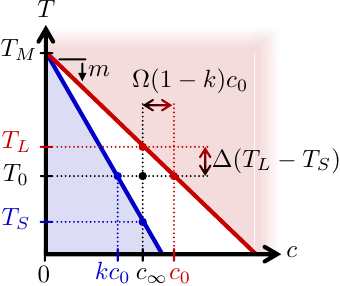}
\caption{
\label{fig:appd1} 
Reference temperatures and concentrations for the solidification of an alloy of nominal concentration $c_\infty$ at a temperature $T_0$.
}
\end{figure}

We use a linear approximation of a binary alloy phase diagram with a liquidus slope $m$ and a partition coefficient $k$ between solid and liquid solute equilibrium concentrations at the solid-liquid interface, as illustrated in Fig.~\ref{fig:appd1}.
The liquidus temperature of an alloy of nominal concentration $c_\infty$ is thus $T_L = T_M + m c_\infty$, and at a set temperature $T=T_0$, dimensionless undercooling and supersaturation are respectively defined as
\begin{align}
\label{eq:delta}
\Delta &= \frac{T_L-T_0}{m c_\infty \left( 1-1/k\right)} ,\\
\label{eq:omega}
\Omega &=  \frac{c_0-c_\infty}{(1-k)c_0} ,
\end{align}
with notations summarized in Fig.~\ref{fig:appd1}.
Using the expression of the equilibrium concentration at $T_0$, i.e. $c_0=(T_M-T_0)/m$, Eqs~\eqref{eq:delta} and \eqref{eq:omega} can be written as
\begin{align}
\label{eq:appdelta}
\frac{c_0}{c_\infty} &= 1-(1-1/k) \Delta ,\\
\label{eq:appomega}
\frac{c_\infty}{c_0} &= 1-(1-k)\Omega ,
\end{align}
which can be combined to yield the relation between dimensionless undercooling and supersaturation
\begin{align}
\label{eq:omegadelta}
\Omega = \frac{1}{1-k}\left[ 1-\frac{1}{1-(1-1/k)\Delta} \right].
\end{align}

Considering alloy parameters from Table~\ref{tab:alcu}, an Al-Cu alloy of nominal composition $c_\infty=10\,$wt\%\,Cu has a unit undercooling $\Delta T_0=T_L-T_S=m c_\infty \left( 1-1/k\right) = 184.3~$K. 
Thus, the imposed undercooling $T_L-T_0=10~$K corresponds to a dimensionless undercooling $\Delta=0.0543$, and following Eq.~\eqref{eq:omegadelta}, a supersaturation $\Omega  = 0.291$.

\subsection{Theoretical steady state}

For $\Omega = 0.291$, the Ivantsov solution \eqref{eq:Iv2D} gives
\begin{align}
\Pec = R_sV_s/(2D) \approx 0.0415 ,
\end{align}
which is to be combined with the solvability condition \eqref{eq:R2V} to yield theoretical steady-state growth conditions.
The solute capillarity length of the solid-liquid interface at the temperature $T_0$ is 
\begin{align}
d_0 =  \frac{\Gamma_{sl}}{|m|(1-k)c_0} \approx 6.98~{\rm nm}, 
\end{align}
with $c_0=13.3$\,wt\%Cu from Eq.\eqref{eq:appdelta} or \eqref{eq:appomega}.
The tip selection parameter is chosen as $\sigma\approx0.08$, which corresponds to a one-sided 2D parabola with an interfacial energy anisotropy $\epsilon_4=0.012$~\cite{BarbieriLanger89}.
The resulting steady-state tip radius and velocity are
\begin{align}
R_s &=  \frac{d_0}{\sigma\Pec} \approx 2.10~\mu{\rm m}, \\
V_s &=  2\sigma \Pec^2 \frac{D}{d_0} \approx 94.8~\mu{\rm m/s}.
\end{align}

\subsection{Buoyancy}

We estimate the solute expansion coefficient at the liquidus temperature $T_L = 903~$K.
We use experimental values for the density listed in Table~\ref{tab:alcu}~\cite{Plevachuk08} to estimate the liquid densities of 
$\varrho({\rm Al-4wt\%\,Cu})\approx 2.436$~g\,cm$^{-3}$ 
and 
$\varrho({\rm Al-20wt\%\,Cu})\approx 2.698$~g\,cm$^{-3}$ 
at $T=T_L$.
From these values, for $c_\infty=10$wt\%\,Cu we interpolate 
\begin{align}
\left.  \varrho \right|_{T=T_L} \approx 2.53~{\rm g/cm}^{3}
\end{align}
and estimate 
\begin{align}
\left. \frac{\partial \varrho}{\partial c} \right|_{T=T_L} \approx 0.0164\times 10^{-3}~{\rm g/cm}^{3}{\rm /wt\%} .
\end{align}
The solute expansion coefficient is then
\begin{align}
\beta_c = \left. -\frac{\frac{\partial \varrho}{\partial c}}{\varrho} \right|_{T=T_L} \approx 6.46\times 10^{-3}~{\rm wt\%}^{-1},
\end{align}
or in dimensionless form 
\begin{align}
\lambda_c = (1-k)\frac{c_\infty}{k}\beta_c \approx 0.397.
\end{align}
We also use experimental measurements listed in Table~\ref{tab:alcu}~\cite{Plevachuk08} to interpolate the liquid viscosity between 
$\eta({\rm Al-4wt\%\,Cu})\approx 1.486$~mPa\,s 
and 
$\eta({\rm Al-20wt\%\,Cu})\approx 1.603$~mPa\,s 
at $T=T_L$.
The resulting dynamic liquid viscosity of Al-10wt\%\,Cu at $T=T_L$ is 
\begin{align}
\eta\approx 1.53 ~{\rm mPa\,s},
\end{align}
i.e. a kinematic viscosity
\begin{align}
\nu=\frac{\eta}{\varrho} = 0.605\times 10^{-6}~{\rm m}^2{\rm s}^{-1}
\end{align}
or in dimensionless form
\begin{align}
\Sch = \frac{\nu}{D} = 252 .
\end{align}
Finally, the dimensionless value of the Earth gravity $\gr=g_y \,\vec{y}$ with $g_y=-9.81~$m/s$^2$ is
\begin{align}
\tilde g_y = \frac{R_s}{V_s^2} g_y \approx -2294.
\end{align}

\subsection{Numerical parameters}
We use a spatial grid spacing $h=2R_s = 4.20~\mu$m, such that the $(2.56~$mm)$^2$ domain consists of 608$^2$ grid points.
The total solidification time of 10~seconds corresponds to $451 R_s/V_s$.
The thickness of the needles is bounded to $r_{max}=2h=4R_s$, and the radius for the integration of the FIF is $r_i=3h$.
The average sidebranching frequency is set to $L_{sb}=10R_s$ and the amplitude of its fluctuation to $\Delta l_{sb}=5R_s$.
Boundary conditions are set to no-flux for the solute field, i.e. $\nabla U\cdot{\mathbf n}=0$, and no-slip for the velocity field, i.e. $u=v=0$ along the boundaries.
Other numerical parameters are $K_{\Delta t}=0.5$, $\omega_{\rm up}=0.9$, $\omega_{\rm sor}=1.7$, and $\overline r_{\rm sor}=10^{-3}$.

\subsection{Initial conditions}
The domain is initialized with a solute supersaturation $U=\Omega=0.291$, with 36 circular nuclei.
These fourfold symmetry crystal nuclei are composed of four branches of equal length $l_0=2h$ and radius $R_0=2h$.
The initial growth kinetics in the early stage is treated with similar equations as for well-developed dendritic branches.
The location and orientation of the 36 seeds are listed in Table~\ref{tab:init}.
\renewcommand{\arraystretch}{1.}
\begin{table}[!t]
\caption{%
Location and orientation of initial nuclei. The last column corresponds to the label of the grain in Figure~\ref{fig:Polycrystal}.
}%
\center%
\label{tab:init}%
\begin{tabular}{c c c c c}%
\hline
seed & $x_0/L_x$ & $y_0/L_y$ & $\theta$ [$^\circ$] & Fig.~\ref{fig:Polycrystal} \\
\hline
1 & 0.13 & 0.11 & 27.46 \\
2 & 0.31 & 0.09 & 38.70 \\
3 & 0.46 & 0.03 & 44.90 \\
4 & 0.51 & 0.11 & 22.92 \\
5 & 0.75 & 0.06 & 88.29 \\
6 & 0.89 & 0.15 & 12.67 \\
7 & 0.09 & 0.25 & 73.81 & L \\
8 & 0.21 & 0.28 & 10.75 & K \\
9 & 0.38 & 0.27 & 50.91 \\
10 & 0.52 & 0.18 & 79.54 & I \\
11 & 0.68 & 0.26 & 83.21 & H \\
12 & 0.93 & 0.28 & 64.66 \\
13 & 0.04 & 0.47 & 48.86 & A \\
14 & 0.27 & 0.48 & 2.44 & C \\
15 & 0.46 & 0.34 & 52.98 & J \\
16 & 0.60 & 0.44 & 24.16 & F \\
17 & 0.70 & 0.35 & 64.37 & G \\
18 & 0.96 & 0.35 & 2.56 \\
19 & 0.08 & 0.51 & 44.84 & B \\
20 & 0.23 & 0.60 & 44.04 \\
21 & 0.37 & 0.66 & 8.89 & D \\
22 & 0.63 & 0.60 & 16.33 & E \\
23 & 0.81 & 0.51 & 67.54 \\
24 & 0.85 & 0.67 & 55.23 \\
25 & 0.06 & 0.67 & 39.58 \\
26 & 0.27 & 0.73 & 14.25 \\
27 & 0.42 & 0.80 & 59.39 \\
28 & 0.57 & 0.72 & 30.50 \\
29 & 0.70 & 0.69 & 37.55 \\
30 & 0.85 & 0.75 & 87.51 \\
31 & 0.16 & 0.92 & 29.32 \\
32 & 0.32 & 0.99 & 59.11 \\
33 & 0.49 & 0.84 & 32.88 \\
34 & 0.65 & 0.93 & 87.26 \\
35 & 0.74 & 0.88 & 42.79 \\
36 & 0.85 & 0.92 & 49.00 \\
\hline
\end{tabular}
\end{table}

\end{appendix} 
\renewcommand{\thefigure}{\arabic{figure}}
\renewcommand{\thetable}{\arabic{table}}

\bibliographystyle{model3-num-names}

\end{document}